\documentclass[english,aps,pre,preprint,superscriptaddress]{revtex4-1}
\usepackage[english]{babel}
\usepackage{float}
\usepackage[latin9]{inputenc}
\usepackage{lmodern}
\usepackage[T1]{fontenc}
\usepackage{amsmath}
\usepackage{graphicx}
\usepackage{makecell}
\let\revappendix\appendix



\begin{document}

\title{Entropic Lattice Boltzmann Model for Charged Leaky Dielectric Multiphase Fluids in Electrified Jets}

\author{Marco Lauricella}
\email[Corresponding author: ]{m.lauricella@iac.cnr.it}
\affiliation{Istituto per le Applicazioni del Calcolo CNR, Via dei Taurini 19, 00185 Rome, Italy}
\author{Simone Melchionna}
\affiliation{Istituto dei Sistemi Complessi CNR, Consiglio Nazionale delle Ricerche, Dipartimento di Fisica, Universit\`a di Roma Sapienza, P.le A. Moro 2, 00185 Rome, Italy}
\author{Andrea Montessori}
\affiliation{Istituto per le Applicazioni del Calcolo CNR, Via dei Taurini 19, 00185 Rome, Italy}
\affiliation{Department of Engineering, University of Rome \textquotedbl{}Roma Tre,\textquotedbl{} Via della Vasca Navale 79, 00141 Rome, Italy}
\author{Dario Pisignano}
\affiliation{Dipartimento di Fisica, Universit\`a di Pisa, Largo B. Pontecorvo 3, 56127 Pisa, Italy}
\affiliation{NEST, Istituto Nanoscienze-CNR, Piazza S. Silvestro 12, 56127 Pisa, Italy}
\author{Giuseppe Pontrelli}
\affiliation{Istituto per le Applicazioni del Calcolo CNR, Via dei Taurini 19, 00185 Rome, Italy}
\author{Sauro Succi}
\affiliation{Istituto per le Applicazioni del Calcolo CNR, Via dei Taurini 19, 00185 Rome, Italy}
\affiliation{Harvard Institute for Applied Computational Science, Cambridge, Massachusetts, United States}

\begin{abstract}
We present a lattice Boltzmann model for charged leaky dielectric multiphase fluids 
in the context of electrified jet simulations, which are of interest for a number of production 
technologies including electrospinning.
The role of non-linear rheology on the dynamics of electrified jets is considered by exploiting
the Carreau model for pseudoplastic fluids.
We report exploratory simulations of charged droplets at rest and under a constant electric field,
and we provide results for charged jet formation under electrospinning conditions.
\end{abstract}


\maketitle

\section{Introduction}
\label{sec1}

The dynamics of charged leaky dielectric jets present a major interest, both as an outstanding problem
in non-equilibrium thermodynamics, as well as for its numerous applications in science and engineering \cite{yarin2014fundamentals,wendorff2012electrospinning,pisignano2013polymer}. 
In particular, the recent years have witnessed a surge of interest towards the manufacturing of electrospun polymeric nanofibers, mostly on account of their prospective applications, such as tissue engineering, air and water filtration, optoelectronics, drug delivery and regenerative medicine \cite{pisignano2013polymer,frenot2003polymer,huang2003review}. 
As a consequence, several experimental studies have focused on the 
characterization and production of one-dimensional elongated nanostructures 
\cite{li2004electrospinning,greiner2007electrospinning,carroll2008nanofibers, huang2003review,persano2013industrial}.
Electrospun nanofibers are typically produced at laboratory scale via 
the uniaxial stretching of a leaky dielectric jet, which is ejected at a nozzle from an electrified 
charged polymer solution. The charged jet elongates under the effect of an 
external electrostatic field applied between the spinneret and a conductive collector
and eventually undergoes electromechanical (e.g., whipping) instabilities due to various sources
of disturbance, such as mechanical vibrations at the spinneret, hydrodynamic friction
with the surrounding fluid and others \cite{reneker2000bending}.
While such instabilities can be detrimental in some respect, making an accurate position of individual fibres on target substrates very hard, in other experiments they are sought for since they
result in thinner cross sections, hence finer electrospun fibres, as they hit the collector \cite{frenot2003polymer}.
This follows from a plain argument of mass conservation: whipping instabilities generate
longer jets, hence thinner cross sections \cite{feng2003stretching}.

The computational modelling of the electrospinning process is based on two 
main families of techniques: particle methods and Lagrangian fluid methods.
The former is based on the representation of the polymer jet as a discrete collection of
discrete particles (beads) connected via elastic springs with frictional coupling 
(dissipative dashpots) and interacting via long-range Coulomb electrostatics \cite{reneker2000bending,lauricella2015jetspin,lauricella2016dynamic,lauricella2016three}.
The latter, on the other hand, describes the jet as a continuum media, obeying the 
Navier-Stokes equations for a charged fluid in Lagrangian form \cite{ganan1997theory,hohman2001electrospinning,yarin2001bending}.
Both methods are grid-free, hence well suited to describe abrupt changes of the jet morphology
without taxing the grid resolution, as it is the case for Eulerian grid methods.  

In this respect, grid-based methods, such as Lattice Boltzmann methods (LBM), are not 
expected to offer a competitive alternative to the two aforementioned class of methods.
Nevertheless, owing to its efficiency, especially on parallel computers, and its
flexibility towards the inclusion of physical effects beyond single-phase hydrodynamics, it appears
worth exploring the possibility of using LBM also in the framework of electrified fluids and jets.
For instance, in the last decade significant improvements in LBM for modelling microfluidic flows containing electrostatic interactions have been achieved \cite{li2003discrete,kupershtokh2006lattice}, opening new applications of LBM in electrohydrodynamic problems 
\cite{huang2007application,gong2010lattice,medvedev2010lattice,bararnia2013breakup,kupershtokh2014three}.
In particular, LBM was successfully employed to simulate deformations and breakup of conductive vapour bubbles, bubble deformation due to electrostriction, dynamics of drops with different electric permittivity.
All these investigations usually exploit the approach originally introduced by A. L. Kupershtokh and D. A. Medvedev \cite{kupershtokh2006lattice},
where dielectric liquids are assumed with zero free charge density, so that the charge carriers are essentially locally bounded to the material \cite{landau2013electrodynamics}.
Within this assumption, charge carriers are explicitly modelled by a convective transport equation solved by a second LB solver, taking into account the rates of ionization and recombination of charge carriers fluctuating around a local value (distribution) of equilibrium.

In the 1960s, G. I. Taylor provided several considerations for dealing with electrified fluid
in a series of papers \cite{taylor1964disintegration,taylor1966studies,taylor1969electrically}. In particular, Taylor discovered that a moving charged fluids cannot be
considered either as a perfect dielectric or as a perfect conductor.
Instead, the fluid acts as a "leaky dielectric liquid", where 
a non-zero free charge is mainly accumulated on the interface between the charged liquid and
the gaseous phase \cite{saville1997electrohydrodynamics}.
As a consequence, the charge produces electric stresses different from those 
observed in perfect conductors or perfect dielectrics.
Indeed, in the last cases, the charge induces a stress which is perpendicular to the interface,
altering the interface shape to balance the extra stress.
In the electrospinning process, the non-zero electrostatic field tangent to the liquid interface produces
a non-zero tangential stress on the interface which is balanced from the viscous force \cite{hohman2001electrospinning}.

The present work exploits the entropic variant of LBM \cite{karlin1999perfect}.
In particular, the use of the entropic lattice Boltzmann method (ELBM) allows to
extend the LBM application also in presence of intense electrostatic forces, acting on 
charged leaky dielectric liquids, which is of main interest for modelling the electrospinning process.
In this context, the largest part of the charge is modelled to lie along the interface between the liquid and the gaseous phase in similarity with previous works 
\cite{hua2008numerical,li2011lattice}.
Further, the present ELBM is generalized to the case of non-Newtonian flows
with a shear-thinning viscosity in order to account the rheological properties of electrospun jets.
Here, we adopt the entropic approach introduced by S. Ansumali and I. V. Karlin in Ref \cite{ansumali2002single}
in order to preserve locally the second principle (H- theorem)
also in presence of sharp changes in the fluid viscosity and structure.

The paper is organised as follows.
In section II we present the basic features of the LB extension to
the case of charged multiphase fluids.
In sections III, we present results for the case of charged multiphase 
fluids at rest and we report on preliminary results for 
charged multiphase jets under conditions related to electrospinning experiments.

\section{Model}
\label{sec2}

We consider a single species, charged fluid as composed of point-like
particles and neglect correlations stemming from excluded volume interactions.
Following Boltzmann's description, the state of the fluid is
determined by the distribution function $f_{p}(\vec{r},t)$ being the probability
of finding at time $t$ the fluid at position $\vec{r}$ and moving with
discrete velocity $\vec{c}_{p}$, with $p=1,b$ given $b$ the number of lattice directions. 
Here, the velocities $\vec{c}_{p}$ are also viewed as vectors connecting
a lattice site $\vec{r}$ to its lattice neighbours. 
The ELBM equation reads
\begin{equation}
f_{p}(\vec{r}+\vec{c}_{p} \Delta t,t+\delta t)=f_{p}(\vec{r},t)-\alpha \beta (f_p - f^{eq}_p(\rho, \vec{u})) + S_{p}(\vec{r},t),
\label{eq:lattice-b}
\end{equation}
where the product $\alpha \beta$ plays the role of a collision frequency, $S_p$ is a source term (see below), and $f^{eq}_p$ is the Maxwell-Boltzmann distribution computed at density $\rho$ and velocity $\vec{u}$ \cite{chikatamarla2006entropic}.
The macroscopic variables
are given by the density $\rho=\sum_{p}f_{p}$ and the fluid velocity
$\vec{u}=1/\rho\sum_{p}\vec{c}_{p}f$. In the following, we refer to
lattice units where the mesh spacing and timestep $\Delta t$ are conveniently
set to unity. Also, we adopt the so-called D2Q9 scheme, composed by
8 discrete speeds (connecting first and second lattice neighbors)
and one extra null vectors accounting for particles at rest. In this
scheme, Here, the $f^{eq}_p$ are chosen
as a second-order Mach-number expansion
\begin{equation}
f_{p}^{eq}=w_{p}\rho\left[1+\frac{\vec{u}\cdot \vec{c}_{p}}{c^{2}_s}+\frac{(\vec{u}\cdot \vec{c}_{p})^{2}-c^{2}_s u^{2}}{2c^{4}_s}\right]
\end{equation}
where
the $w_p$ are weights equal to 4/9 for the rest particles, 1/9 and 1/36 respectively for the smallest
and largest velocities $\vec{c}_{p}$, and
$c_s$ is the speed of sound that in lattice units is equal
to $1/\sqrt{3}$. At the same time, we consider a unit fluid molecular
mass, so that the thermal energy is equal to $k_B T=c^{2}_s$ with $k_B$ the Boltzmann constant and $T$ the temperature.
Following the approach of Refs. \cite{dorschner2016entropic,chikatamarla2006entropic,ansumali2002single}, the factor $\beta$ in Eq. \ref{eq:lattice-b} depends on the kinematic viscosity $\nu$ by the relation
\begin{equation}
\beta=\frac{c^2_s}{2\nu+c^2_s},
\label{eq:beta}
\end{equation}
while $\alpha$ is the largest value of the over-relaxation parameter so that
the local entropy reduction can be avoided, ensuring the H- theorem.
In particular, $\alpha$ is computed as the root of the scalar nonlinear equation \cite{ansumali2002single,karlin1999perfect} 
\begin{equation}
H\left(f+\alpha (f^{eq}-f) \right)=H\left(f \right),
\label{eq:entropic}
\end{equation}
where $H$ denotes the Boltzmann's entropy function, defined in discrete form \cite{ansumali2003minimal} as
\begin{equation}
H\left(f \right) \equiv \sum_p f_{p} \ln \left(\frac{f_{p}}{w_{p}} \right).
\label{eq:H}
\end{equation}

In Eq. \ref{eq:lattice-b}, the source term $S_p$ takes into account the global effect of all the
internal and external forces $\vec{F}$. This is assessed by the exact difference method 
proposed by Kupershtokh et al. \cite{kupershtokh2009equations}, which reads
\begin{equation}
S_{p}= f_{p}^{eq} \left(\rho,\vec{u} + \Delta \vec{u} \right)-f_{p}^{eq} \left(\rho,\vec{u} \right),
\label{eq:add}
\end{equation}
where $\Delta \vec{u}=\vec{F} /\rho$.
In bulk conditions, the ELBM is intrinsically second-order accurate
in space and time, and, in order to ensure the same accuracy in presence
of forces, the local velocity is taken at half time step
\begin{equation}
\rho \vec{u}=\sum_{p}f_{p}(\vec{r},t)\vec{c}_{p} + \frac{1}{2}\vec{F}.
\end{equation}
 
Here, the total body force $\vec{F}=\vec{F}_{int}+\vec{F}_{el}$ includes the inter-particle force $\vec{F}_{int}$ and the electric force $\vec{F}_{el}$. 
The electric force acting on the boundary point $\vec{r}$ between a gas and a fluid with the local non-uniform permittivity $\varepsilon(\vec{r})$ 
in an electric field $\vec{E}$ reads \cite{landau2013electrodynamics,saville1997electrohydrodynamics}
\begin{equation}
\begin{aligned}
\vec{F}_{el} &= q \vec{E} -\frac{1}{2} |E|^2 \nabla \varepsilon +\frac{1}{2} \nabla \left(|E|^2 \rho \frac{\partial \varepsilon}{\partial \rho} \right) \\
&= q \vec{E} + \frac{1}{2} \rho \frac{\partial \varepsilon}{\partial \rho} \nabla |E|^2  ,
\end{aligned}
\label{eq:force_el}
\end{equation}
where $q$ is the local free charge carried on the fluid.
In the last Eq., the vacuum permittivity $\varepsilon_0$ was assumed equal to 1 as in the Gaussian centimetre-gram-second (cgs) unit system, so that the charge in lattice units is
$length^{3/2} \, mass^{1/2} \,time^{-1}$ in similarity with the statcoulomb definition (note that Coulomb's constant is also 1). For the sake of convenience, we report in Appendix \ref{app:table} the units conversion Table \ref{tab:unit} in cgs
dimensions from lattice units for several physical quantities shown in the following.
As in Ref. \cite{kupershtokh2006lattice}, we consider a fluid with permittivity $\varepsilon=1+\rho/\rho_0$
with $\rho_0$ an arbitrary constant (in the following taken for simplicity equal to 1) so that it is
$\rho(\partial \varepsilon / \partial \rho)=\varepsilon-1$.
As a consequence, Eq. \ref{eq:force_el} reduces to
\begin{equation}
\vec{F}_{el} = q \vec{E} + \frac{ \varepsilon-1 }{2} \nabla  |E|^2 .
\label{eq:force_el2}
\end{equation}

In the following, we assume that the magnetic induction effects can be neglected so $\nabla \times \vec{E} = 0$, 
and the system follows the Gauss law  $\nabla \cdot (\varepsilon \vec{E})=q$.
Since $\vec{E}=-\nabla \phi$  with $\phi$ the electric potential, the Poisson equation $div(\varepsilon(\vec{r}) \nabla \phi )=-q(\vec{r})$ can be solved at each lattice node $\vec{r}$, given the boundary conditions of the system and the local
charge $q(\vec{r})$ at the node (specified below).
In particular, we determine the electric potential by solving numerically the two-dimensional
Poisson equation by means of a SOR (Successive Over-Relaxation) algorithm
and the Gauss-Seidel method \cite{quarteroni2008numerical}.
Note that the Poisson equation includes the non-uniformity of the permittivity $\varepsilon(\vec{r})$, and it is solved on-the-fly
during the simulation. Hence, the electric force $\vec{F}_{el}=-q \nabla \phi$ is added
into the ELBM by Eq. \ref{eq:add}. 

Since we are modelling a leaky dielectric fluid, we assume that the free charge in the system is mainly distributed over the liquid-gaseous interface. 
Further, in similarity to previous electrospinning models \cite{hohman2001electrospinning,yarin2001bending}, 
the relaxation time of free charge in the system is assumed to be irrelevant.
In other words, the free charge in bulk liquid relaxes to the liquid interface in a smaller time than any other characteristic time
in the system \cite{ganan2004general}. 
This is a well-established assumption of a leaky dielectric fluid (for further details see Ref. \cite{saville1997electrohydrodynamics}).
The liquid charge in the point $\vec{r}$ is given as 
\begin{equation}
q=q_b+q_s,
\label{eq:sum_q}
\end{equation}
which is the sum
of a surface charge $q_s$ and a small bulk term $q_b$. 
The bulk term $q_b$ is taken as 
\begin{equation}
q_b(\vec{r})=Q_b \frac{ \rho(\vec{r}) \, \theta(\rho(\vec{r});\rho_0)}{\int \rho(\vec{r}) \, \theta(\rho(\vec{r});\rho_0) d\vec{r}},
\end{equation}
where $Q_b$ denotes the total charge in the bulk, the denominator acts to keep constant
the charge due to the charge conservation principle, and
$\theta(\rho;\rho_0)$ denotes a smoothed version of the Heaviside step function switching
from zero to one at $\rho_0$ (equal to 1 in all the following simulations) in order to select only the liquid phase.
The term $q_s$ is modeled as a proportional to the absolute density gradient
\begin{equation}
q_s(\vec{r})= Q_s \frac{|\nabla \rho(\vec{r})|^2}{\int |\nabla \rho(\vec{r})|^2 d\vec{r}},
\label{eq:surface-q}
\end{equation}
where $Q_s$ denotes the total charge over the surface, and the denominator ensures the charge conservation principle as in the previous case.
This approach is usually referred to as the constant surface charge model. It is important to highlight that such charge model is different from the method adopted
by Kupershtokh et al. \cite{kupershtokh2006lattice}, where the charge carriers are treated by using
an additional LB component with zero mass (passive scalar) to model the polarizability of a dielectric liquid with zero free charge.
Since in the present model the charge is directly modelled over the interface, we do not need to introduce any extra LB component to model the surface charge.
Indeed, the constant surface charge model was already adopted in Refs \cite{hua2008numerical,li2011lattice} as a strategy to simplify the  charge transport and 
distribution on the droplet interface. Nonetheless,  the constant surface charge model fails in describing
a distributed charge on the drop interface
whenever the charge density is high, since the curvature 
surface alters the local charge density \cite{hohman2001electrospinning}.
In order to address the issue, we assume that the curvature biases the surface charge density
as in a conductive liquid, following the power-law introduced by I. W. McAllister \cite{mcallister1990conductor}, which states
\begin{equation}
q_s=q_{s,max}  (K/K_{max})^{\frac{1}{4}}.
\label{eq:surface-curvature}
\end{equation}
Here, $K$ denotes the mean curvature $ K=\nabla \cdot \widehat{n}$ with the local interface normal
$\widehat{n}=\nabla \rho(\vec{r})/ |\nabla \rho(\vec{r})|$ \cite{spencer2010lattice}, while $q_{s,max}$ is the maximum surface charge at the
maximum curvature $K_{max}$ chosen as a reference value for the system under investigation. It is worth to 
emphasize that treating a leaky dielectric as a conductive liquid is a simplification already made
by several authors (e.g., G. Taylor \cite{taylor1964disintegration}, A. Yarin et al. \cite{yarin2001taylor} , etc.). For the sake of simplicity,
we take in the following the maximum curvature $K_{max}$ equal to $K_{d}$ value, defined as the curvature
doubling the local surface charge density.
Thus, we rewrite Eq. \ref{eq:sum_q} as
\begin{multline}
q(\vec{r})=Q_b \frac{ \rho(\vec{r}) \, \theta(\rho(\vec{r});\rho_0)}{\int \rho(\vec{r}) \, \theta(\rho(\vec{r});\rho_0) d\vec{r}}+   \\
+ Q_s \frac{|\nabla \rho(\vec{r})|^2 [1+(K/K_{d})^{\frac{1}{4}}]}{\int |\nabla \rho(\vec{r})|^2 [1+(K/K_{d})^{\frac{1}{4}}] d\vec{r}}.
\label{eq:surface-q2}
\end{multline}
The total charge of the system is conserved and equal to $Q=Q_b + Q_s$.

In addition, the fluid is subjected to an internal thermodynamic
force $\vec{F}_{int}$ promoting a phase separation, in similarity with the approach originally 
introduced by Mazloomi et al. \cite{chikatamarla2015entropic} in the context of the ELBM.
The phase separation force is accounted for by means of the Shan-Chen method \cite{huang2015multiphase,falcucci2010lattice}. We construct the local force as
\begin{multline}
\vec{F}_{int}(\vec{r},t)=-  G \, \psi(\rho(\vec{r},t)) \sum_{p\in fluid}w_{p}\psi(\rho(\vec{r}+ \vec{c}_{p},t))\vec{c}_{p}-    \\ 
-G_{w} \,  \psi(\rho(\vec{r},t)) \sum_{p\in wall}w_{p} \psi(\rho(\vec{r},t)) \vec{c}_{p},
\label{eq:sc}
\end{multline}
with the sum $\sum_{p\in fluid}$ running over lattice nodes where the fluid is allowed,
that is, not belonging to the wall, and $\sum_{p\in wall}$ runs over nodes belonging to
the wall. $G$ and $G_{w}$ are fluid-fluid and fluid-wall interaction strengths, respectively. 
In Eq \ref{eq:sc}, $\psi$ is an effective number density, which is taken for simplicity  
$\psi(\rho)=\rho_0 [1- \exp(-\rho / \rho_{0})]$, being $\rho_{0}$ an arbitrary constant \cite{shan1993lattice}
(in the following assumed equal to 1).

\subsection{Extension to Non-Newtonian flows}
\label{subsec2}

In the electrospinning process, the rheological behaviour 
of polymeric liquid with shear-rate-dependent viscosity is expected to play a significant role in jet dynamics.
As a consequence, we now generalize the present model to the case of non-Newtonian flows, 
in similarity with the
approach reported in Refs \cite{pontrelli2009unstructured,gabbanelli2005lattice,aharonov1993non}.
The shear-rate $\dot{\gamma}$ is a functional of the density distribution function $f$. In particular,
the strain tensor $\mathbf{\Gamma}_{\eta,\delta}$ reads \cite{pontrelli2009unstructured,ouared2005lattice}
\begin{equation}
\mathbf{\Gamma}_{\eta,\delta}=-\frac{1}{2\rho \tau c^2_s}\mathbf{\Pi}_{\eta,\delta},
\label{eq:strain}
\end{equation}
where
\begin{equation}
\mathbf{\Pi}_{\eta,\delta}=\sum_{p}\left(f_p - f_p^{eq} \right)\vec{c}_{p \eta} \vec{c}_{p \delta},
\label{eq:stress}
\end{equation}
is the the stress tensor with $\eta$ and $\delta$ running over the spatial dimensions.
Note that $\tau$ in Eq. \ref{eq:strain} is defined as the inverse of  the product $\alpha\beta$, where $\alpha$ was computed by Eq. \ref{eq:entropic}, and $\beta$ depends by Eq. \ref{eq:beta} on the kinematic viscosity $\nu$.
We now rewrite Eq. \ref{eq:strain} as
\begin{equation}
\dot{\gamma}=\frac{\Pi}{\rho \tau(\dot{\gamma}) c^2_s} ,
\label{eq:strain2}
\end{equation}
where $\dot{\gamma}$ and $\Pi$ are computed as matrix 2-norm $\dot{\gamma}=2||\mathbf{\Gamma}||_2$ and
$\Pi=||\mathbf{\Pi}||_2$ of the shear and stress tensor \cite{pontrelli2009unstructured}, respectively.
Note that in the last Eq. we exploit a constitutive relation between the kinematic viscosity $\nu$ and the shear-rate $\dot{\gamma}$,
so that $\tau=\tau(\dot{\gamma})$. As a consequence, $\dot{\gamma}$ is computed as the root of the scalar nonlinear Eq. \ref{eq:strain2}.

We should now consider the general trend observed in
electrospun polymeric filaments \cite{agarwal2016electrospinning}.
As main features, we highlight that a polymeric spinning solution at low shear rate  
behaves as a quasi-Newtonian fluid with zero shear kinematic viscosity $\nu_0$, since the initial condition can be recovered,
while at a high shear rate a non-reversible disentanglement is present.
In particular, it is possible to identify a relaxation time $\lambda$ at which the shear-thinning starts,
which is equal to the inverse value of the shear rate at that instant.
At the very high shear rate, a quasi-Newtonian behaviour is again observed as soon as the alignment
of the polymer chains is extremely high (almost complete). The last region is characterized by a final
viscosity value (infinite viscosity $\nu_{\infty}$), which is lower than $\nu_0$.

In the present investigation, we exploit the Carreau model \cite{johnson2016handbook},
which is able to describe all the mentioned rheological properties.
The Carreau model states that
\begin{equation}
\nu(\dot{\gamma})=\nu_{\infty}+\left(\nu_{0}-\nu_{\infty} \right) \left[ 1 + \left(\lambda \dot{\gamma} \right)^2 \right]^{(n-1)/2} ,
\label{eq:carreau}
\end{equation}
where $n$ is the flow index ($n<1$ for a pseudoplastic fluid).
Obtained $\dot{\gamma}$ by resolving Eq. \ref{eq:strain2} and $\nu(\dot{\gamma})$ by Eq. \ref{eq:carreau}, and assuming a slow variation of $\nu(\dot{\gamma})$ over the time $\Delta t$, the local parameter $\beta$
is finally estimated by Eq. \ref{eq:beta}.
Note that a validation of a similar implementation in a LB scheme of the Carreau model was given in Ref. \cite{pontrelli2009unstructured}.

\subsection{Summary of the model}

To solve the Eq. \ref{eq:lattice-b}, we exploit the method of
splitting the model procedure into physical process stages. 
Hence, the time step is given by the sequential implementation of the following key points:
\begin{enumerate}
\item Compute the Boltzmann's entropy function, and determine $\alpha$ by solving Eq. \ref{eq:entropic}.
\item Solve the nonlinear Eq. \ref{eq:strain2} to compute the local rate strain $\dot{\gamma}$, and determine the local viscosity $\nu$ as a function of the rheological Eq. \ref{eq:carreau}. Hence, determine $\beta$ from Eq. \ref{eq:beta}.
\item Compute the local charge $q$ by Eq. \ref{eq:surface-q2}, and solve the Poisson equation for assessing the electric force $\vec{F}_{el}=-q \nabla \phi$.
\item Compute the phase separation force $\vec{F}_{int}$ by Eq. \ref{eq:sc}.
\item Apply the Eq. \ref{eq:lattice-b}.
\end{enumerate}

\section{Numerical results}
\label{sec3}

\subsection{Charged drop}
In order to assess the properties of our implementation for charged multiphase systems,
we have initially run a set of simulations modelling a 
charged leaky dielectric fluid system obeying the Shan-Chen equation of state \cite{shan1993lattice}. 
In order to assess the static behaviour,
we take a two-dimensional periodic mesh made of 320 x 320 nodes, 
and prepare the system by creating a circular drop of density $\rho=2.0$ and radius $R=40$ in lattice units,
immersed in the second background phase at lower density $\rho_b=0.16$. 
Further, the strength of non-ideal interactions was set equal to $G = -5$, 
$G/G^o_{crit} = 1.25 $ where $G^o_{crit}=-4$ is the critical Shan-Chen coupling at the critical
density $\rho_{crit}= \ln 2$ in the absence of
electric fields.
Since we aim to model a leaky dielectric fluid, the ratio $Q_s/Q_b$ is taken equal to 10,
so that the largest part of the charge lies over the surface. The total charge $Q=Q_s+Q_b$ was set
equal to $2.13$, and $q(\vec{r})$ 
computed by Eq. \ref{eq:surface-q2}. Whenever the Poisson equation is solved, a uniform negative charge is added to obtain a system with net charge zero.
Hence, the electric force $\vec{F}_{el}=q(-\nabla{\phi})$ is added into the ELBM by Eq. \ref{eq:add},
where $q$ is computed by Eq. \ref{eq:surface-q2} with $K_d$ equal to 1. 
The liquid is Newtonian with kinematic viscosity $\nu=1/6$.

\begin{figure}
\begin{centering}
\includegraphics[width=0.9\columnwidth]{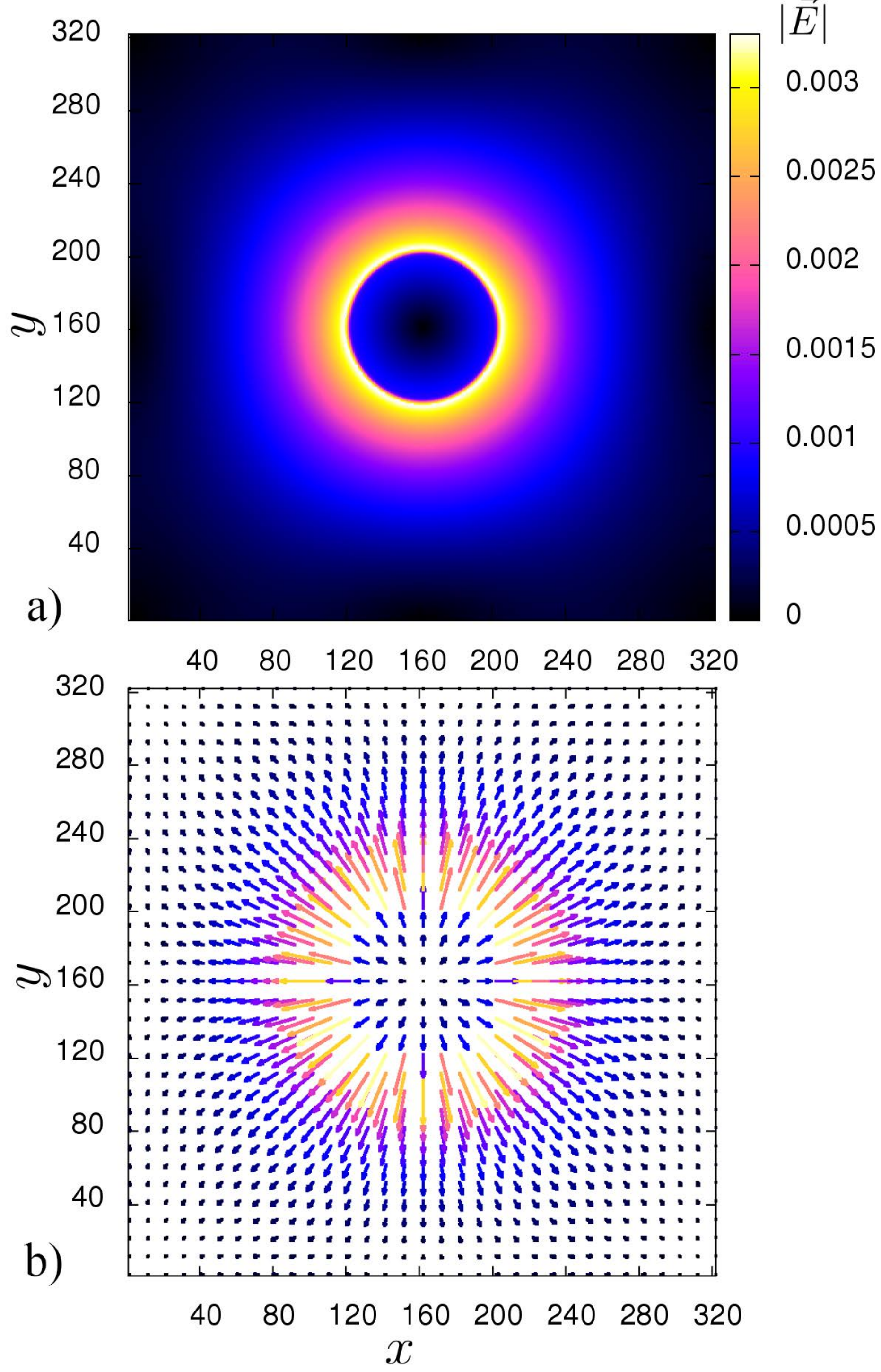}

\par\end{centering}

\protect\caption{Profile of the electric field magnitude $|\vec{E}|$ (panel a), 
and its vectorial representation (panel b). Both quantities refer to the charged drop at rest.}

\label{fig:01}
\end{figure}

The stationary configuration of the described system is obtained after 1000 time steps.
Hence, we inspect the electric field (see Fig. \ref{fig:01}) at rest conditions
in order to analyze the balance of forces acting at the interface, 
including Shan-Chen pressure and capillary and electrostatic forces, the latter
pointing normal to the interface (see panel b of Fig. \ref{fig:01}).
Here, we observe that the largest part of the electric field is located 
over the liquid surface where the charge distribution is higher.
In particular, at the boundary of the drop the magnitude of
the electric field $|\vec{E}|$ is equal to $3.5 \cdot 10 ^{-3}$.

It is of interest to estimate the various forces which concur to provide a stable
configuration of the droplet.
The mechanical balance reads as follows: 
\begin{equation}
p_L + p_{el} =  p_V + p_{cap}
\end{equation} 
where $p_L$ and $p_V$ are the liquid and vapour pressure, respectively, $p_{cap}=\sigma/R$
is the capillary pressure (given the surface tension $\sigma$ and the drop radius $R$), and $p_{el}$ is the repulsive electrostatic pressure.
The latter can be estimated by standard considerations in electrostatics, namely: 
\begin{equation}
p_{el} = \frac{Q_s \vec{E_s}  \cdot \vec{n}}{2 \pi R} 
\end{equation} 
where $E_s$ is the electric field at the surface.
\begin{equation}
\frac{p_{el}}{p_{cap}} = \frac{Q_s E_s}{2\pi \sigma}
\end{equation} 
In actual numbers with $\sigma=5.8 \cdot 10^{-2}$, this ratio is equal to $1/50$.
This shows that electrostatic forces act as a small perturbation on top of the
neutral multiphase physics.

\begin{figure}
\begin{centering}
\includegraphics[width=0.95\columnwidth]{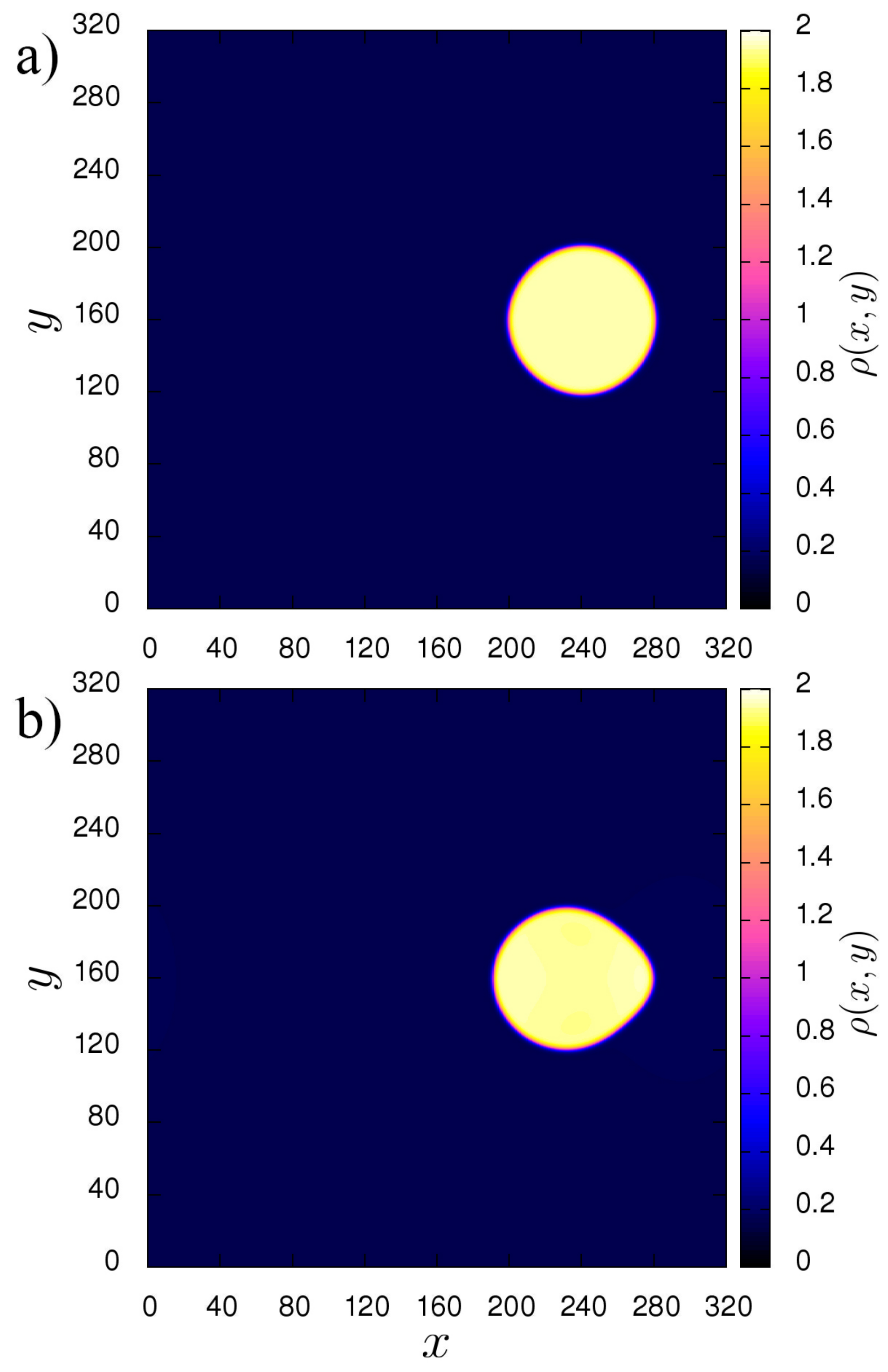}

\par\end{centering}

\protect\caption{Two snapshots of density $\rho$ for a charged drop under an external 
electric field $E_{ext}$ equal to $0.1$ (panel a) and $0.5$ (panel b) taken as soon as the
drop reaches the point of coordinates $(280,160)$.}

\label{fig:02}
\end{figure}

Next, we investigate the effects of a uniform external electric field $E_{ext}$ of magnitude pointing along the $x$ axis. 
Using the previous configuration at the equilibrium 
as the starting point of our simulation, we set $E_{ext}$ at two different values equal to 0.1 and $0.5$.
For each one of the two cases, we report a snapshot of the fluid density $\rho$
taken as soon as the liquid drop touches the point of coordinates (280,160) in lattice units.
The set in Fig. \ref{fig:02} highlights the significant motion of the charged drop to the right in accordance with 
the direction of the electric field. 
In the figure, we note that a sizable change in the drop shape is present only for the case $E_{ext}=0.5$.
In order to elucidate this effect, it is instructive to assess the strength of the electrostatic field 
in units of capillary forces, namely:
\begin{equation}
\tilde E \equiv \frac{Q E_{ext}}{2 \pi R} \frac{R}{\sigma}= \frac{Q E_{ext}}{2 \pi \sigma} 
\end{equation}
In actual numbers, this ratio is equal to 0.5 and 3 for the case at lower and higher $E_{ext}$, respectively.
This shows that the electric force magnitude is sufficiently large to provide an alteration of its shape
only in the case at higher $E_{ext}$. In particular, the shape shows an elongation towards the direction of the electric 
field $E_{ext}$, which results from the effect of the curvature on the surface charge. 

\begin{figure}
\begin{centering}
\includegraphics[width=0.95\columnwidth]{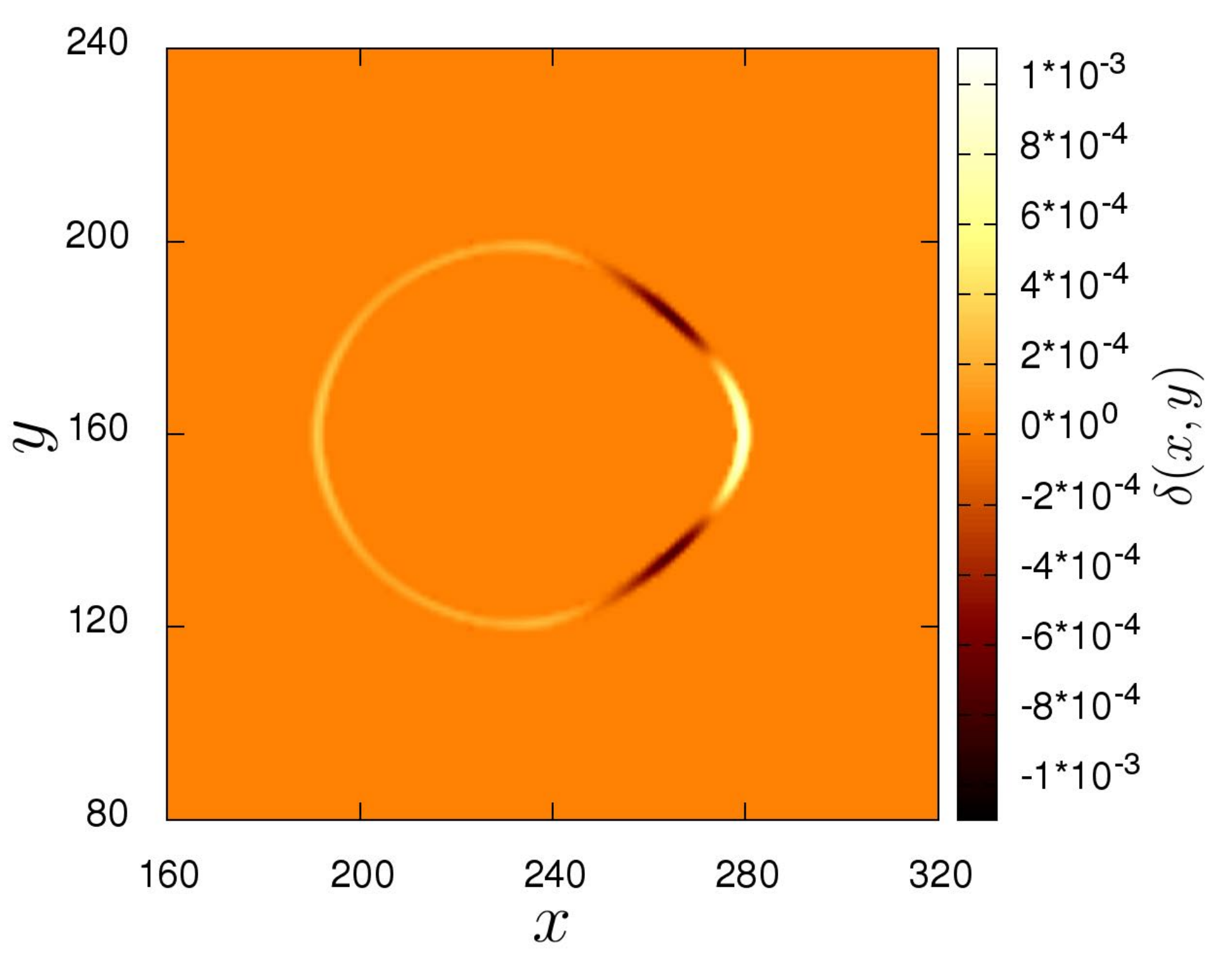}

\par\end{centering}

\protect\caption{Alteration of charge density $q$ due to the mean curvature $K$. The alteration is estimated as $\delta_q=q-q_{K=0}$,
where $q_{K=0}$ is computed with Eq. \ref{eq:surface-q2} with $K=0$ everywhere.}
\label{fig:03}
\end{figure}

\begin{figure}
\begin{centering}
\includegraphics[width=0.95\columnwidth]{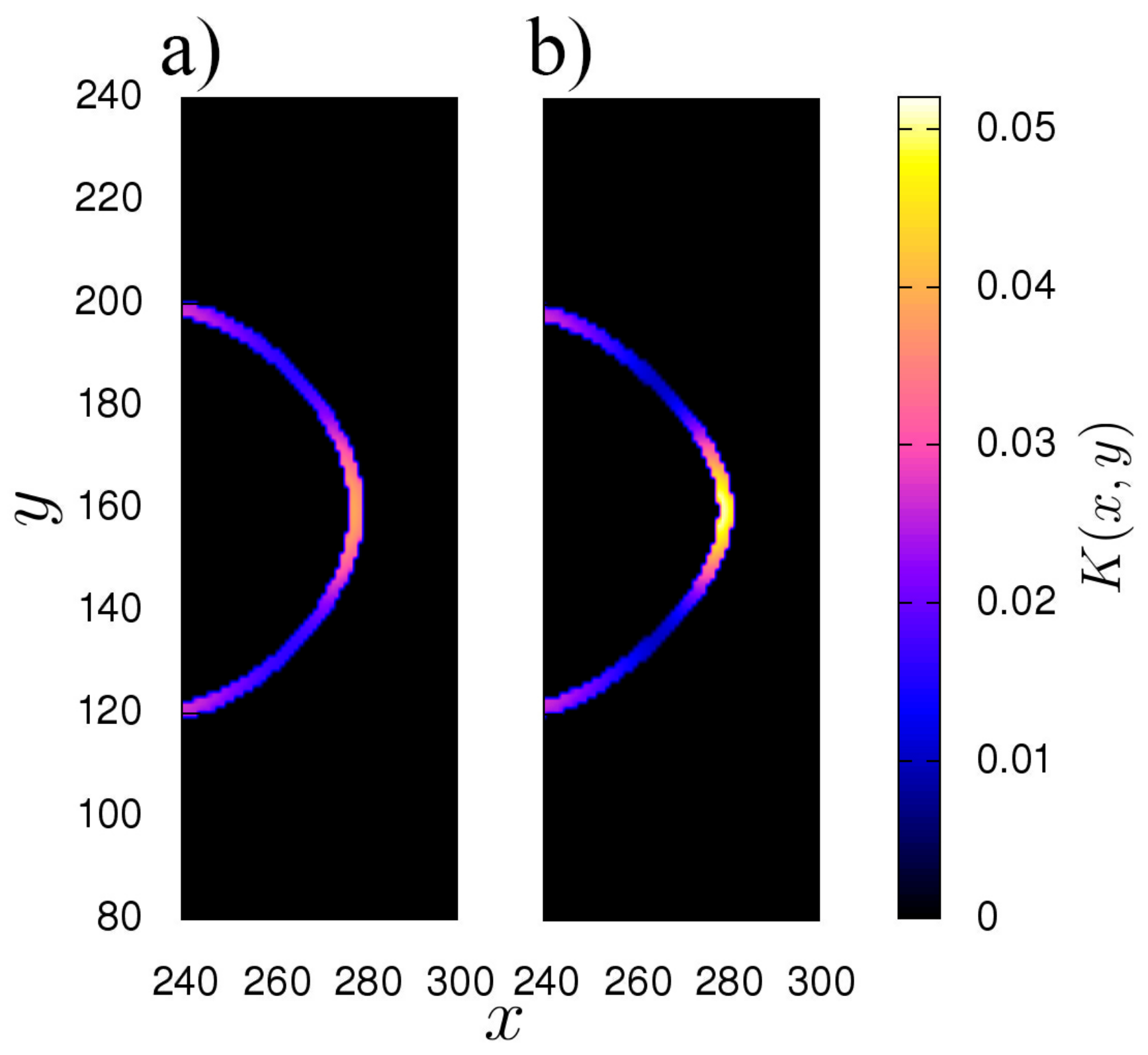}

\par\end{centering}

\protect\caption{Mean curvature $K(x,y,)$ computed at the same time $t=1250 \; \delta t$ with same surface charge $Q_s$ at $E_{ext}=0.5$ with two constant surface charge models:
without the curvature effect (panel a), and with the curvature effect (panel b). }
\label{fig:04}
\end{figure}

In Fig. \ref{fig:03}, we report the alteration of charge density due to the mean curvature term $K$ of Eq. \ref{eq:surface-q2}.
The alteration is estimated as $\delta_q=q-q_{K=0}$, where $q_{K=0}$ is computed with Eq. \ref{eq:surface-q2} with $K=0$ everywhere, corresponding to
a constant surface charge model without curvature effect correction.
Here, we note an accumulation of charge on the rightest part of the drop, where the mean curvature $K$ shows a maximum value equal to 
$5.17 \cdot 10^{-2}$, which corresponds to a charge accumulation $\delta_q$ equal to $1 \cdot 10^{-3}$, in the following denoted $\delta_q^{+}$. The accumulation of charge is counterbalanced by a negative charge $\delta_q^{-}$ distributed over the almost straight surface part of the drop (just behind the rightest protrusion in Fig. \ref{fig:03}). Both partial charges $\delta_q^{+}$ and $\delta_q^{-}$ favour the presence of a protrusion in the drop shape.
In order to analyze this effect, we report in Fig. \ref{fig:04} the mean curvature computed at the same time $t=1250 \delta t$ for
two simulations, both at $E_{ext}=0.5$ differing for the inclusion of the curvature effects in the constant surface charge model of Eq. \ref{eq:surface-q2}.
Even though the circular shape of the drop is deformed by the external electric field in both cases, the curvature effects increase the protrusion on the drop shape
(see Fig. \ref{fig:04} panel b). Further, the charge differences provide a shift in the electric force acting on the drop surface,
the effect of which accumulates in time, so that the alteration in the drop shape increases in time.

\subsection{Charged multiphase jet in electrospinning setup}
\label{sec4}

\begin{figure}
\begin{centering}
\includegraphics[width=0.95\columnwidth]{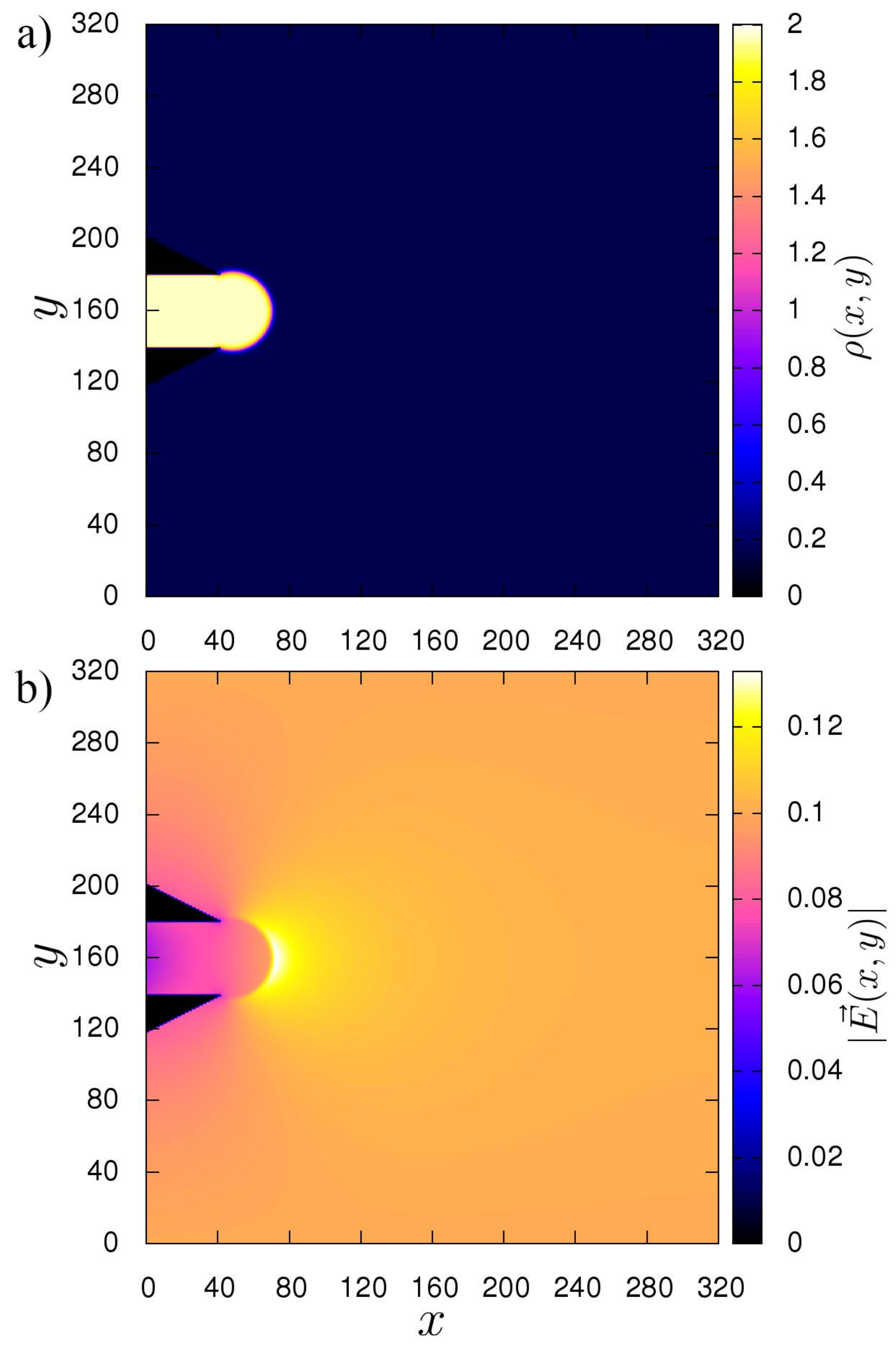}

\par\end{centering}

\protect\caption{In panel a, the density distribution $\rho$ of the initial configuration for the electrospinning setup.
In panel b, the electric field magnitude $|\vec{E}|$. }

\label{fig:05}
\end{figure}

We set up a system modelling the electrospinning process, containing a charged Shan-Chen fluid. 
The system is a mesh made of 320 x 320 nodes (see Fig. \ref{fig:05}). The system geometry presents, on the left side, 
a nozzle of diameter $D=40$, providing an initial jet radius $a_0=20$, that reproduces the needle of the actual electrospinning apparatus 
where the charged fluid is injected, while on the up and bottom sides we impose the bounce back boundary condition.
As a consequence, the system is open with the inlet nodes located inside (left side) 
the nozzle (at $x=1$).
Similarly, we set outlet nodes on the right side (at $x=320$) 
where the jet will impinge under the effect of the external electric field.
Such electric field  is chosen to mimic the potential
difference that is normally applied between the nozzle and a conductive collector 
in the real electrospinning setup \cite{montinaro2015sub}.
The computational setup is quite sensitive to the choice of the simulation parameters, 
and numerical stability has to be guaranteed by finely tuning several parameters,
in particular: the density and velocity of inlet and outlet nodes,
the Shan-Chen coupling constants of fluid-fluid and fluid-wall interactions, the charge constant and
the magnitude of the external electric field. 
After preliminary simulations, we obtain a consistent set of parameters that guarantees a stable and
well-shaped charged jet ejected from the nozzle. 

The initial density of the two phases is $2.0$ and $0.16$ for the liquid and gaseous phase, respectively. 
The initial configuration consists of the liquid phase filling the inner space of nozzle 
with a liquid drop just outside the needle (see Fig. \ref{fig:05} panel a).
All remaining fluid nodes are initialized to gaseous density.

Both the Shan-Chen constants for the fluid-fluid $G$ and fluid-wall $G_w$ interaction are set to $-5$.
As in the previous section, for the resolution of the Poisson equation, a uniform negative charge is added to the system in order to counterbalance the positive charge and obtain a system with net charge zero.
Further, we impose Dirichlet boundary condition in the following form:  we impose the electric potential $\phi_l=0$ on the left side ($x=0$), 
while the electric potential on the right side ($x=321$) was set equal to $\phi_r=-32.2$, 
providing a background electric field $E_{back}=(\phi_r-\phi_l)/322$, which is
imposed between the two opposite sides (left-right) of the system.
On the upper and bottom sides, the Dirichlet boundary conditions are set equal to the
$\phi_u(x)=\phi_b(x)=x(\phi_r-\phi_l)/322$. 
Note that the last condition is equivalent to
impose an electric field $\vec{E}$ of magnitude $0.1$ oriented along the x-axis.

Since in a typical electrospinning setup the liquid phase is always connected to a generator addressing a charge,
it is reasonable to assume that, whenever the stretching of the liquid jet increases the jet interface, extra charge rapidly reaches
the liquid boundary in order preserve the value of charge surface density. 
As a consequence, the charge conservation condition cannot be applied (the liquid jet is not insulated).
Instead, we assume the conservation condition of the surface charge density value for the same mean curvature,
so that Eq. \ref{eq:surface-q2} is rewritten as
\begin{equation}
q(\vec{r})=\xi_b \rho(\vec{r}) \, \theta(\rho(\vec{r});\rho_0)+ \xi_s |\nabla \rho(\vec{r})|^2 [1+(K/K_{d})^{\frac{1}{4}}],
\label{eq:surface-q3}
\end{equation}
where we have adopted a similar condition also for the bulk charge term, being $\xi_b$ and $\xi_s$ two
proportionality constants, in the following taken equal to $1\cdot 10^{-4}$ and $6\cdot 10^{-2}$, respectively. 
Note that the two proportionality constants were tuned in order to obtain a 
mean ratio $Q_s/Q_b$ between the surface and bulk charge close to the target value 10, as in the previous case.

At the inlet, we set the fluid velocity
in accordance with the Poiseuille velocity profile, while the density is set to $2.0$. 
In particular, at each time step, we compute the mean velocity of the fluid
inside the nozzle, then we used this value to set up the Poiseuille profile.
As a consequence, the velocities at the inlet nodes are not fixed but can change 
during the simulation according to the actual mean velocity measured inside the nozzle. 
The outlet nodes (on the right edge) are put in contact with a gas reservoir with $\rho=0.16$, so that the liquid exits
by diffusion/advection.

We run three different simulations, all starting from the same initial configuration. In the first simulation the 
liquid is Newtonian with kinematic viscosity $\nu=1/6$, in the following denoted $case\;1$.
In the other two, we employ the Carreau model (see Subsec. \ref{subsec2}) with zero shear kinematic viscosity 
$\nu_0=1/6$, and infinite viscosity $\nu_{\infty}=0.001$.
The flow index $n$ is taken equal to 0.75, and 0.5, for the case labeled $b$, and $c$, respectively,
while the relaxation time $\lambda$ was set equal to $1000$ for all the two last cases.

The internal electric field computed by the Poisson solver (see Section 2) is computed on-the-fly during the simulation.  In panel b of Fig. \ref{fig:05} we report the electric field magnitude $|\vec{E}|$ for the initial configuration.
Here, we note a maximum value of $|\vec{E}|$ close to the drop interface, which is due to the higher surface charge density.
Further, a lower value of $|\vec{E}|$ is observed in the nozzle as a consequence of the larger 
dielectric constant $\varepsilon \simeq 3$ in the liquid phase (versus $\varepsilon \simeq 1$ in the gaseous phase).

\begin{figure}
\begin{centering}
\includegraphics[width=0.95\columnwidth]{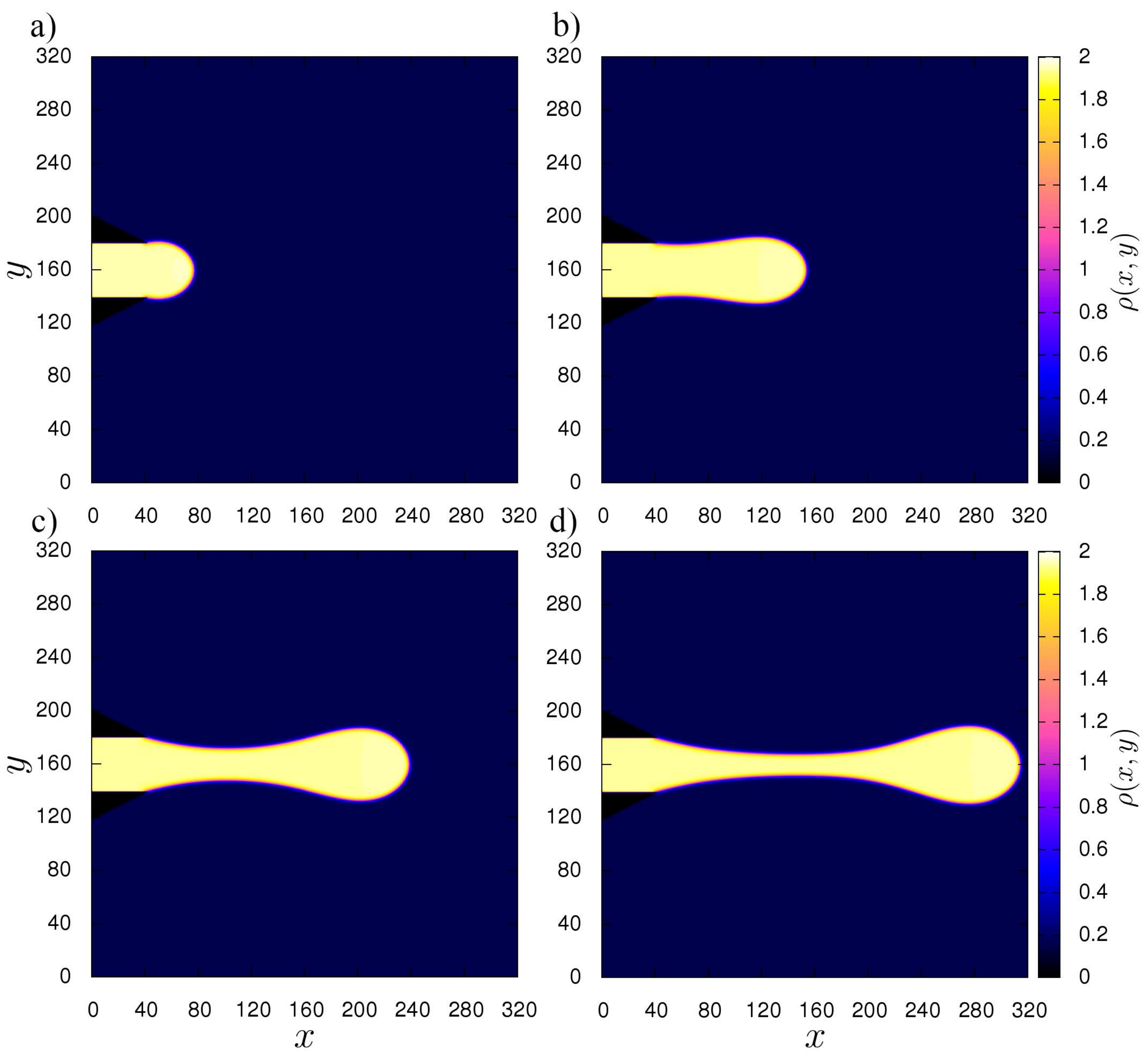}

\par\end{centering}

\protect\caption{Series of snapshots of the fluid density $\rho$
for the electrospinning simulation $case \; b$ with index flow $n=0.75$ and relaxation time $\lambda = 1000$ taken at timesteps 600(a), 5000(b), 7500(d), and 8800(d).}

\label{fig:06}
\end{figure}

\begin{figure}
\begin{centering}
\includegraphics[width=0.95\columnwidth]{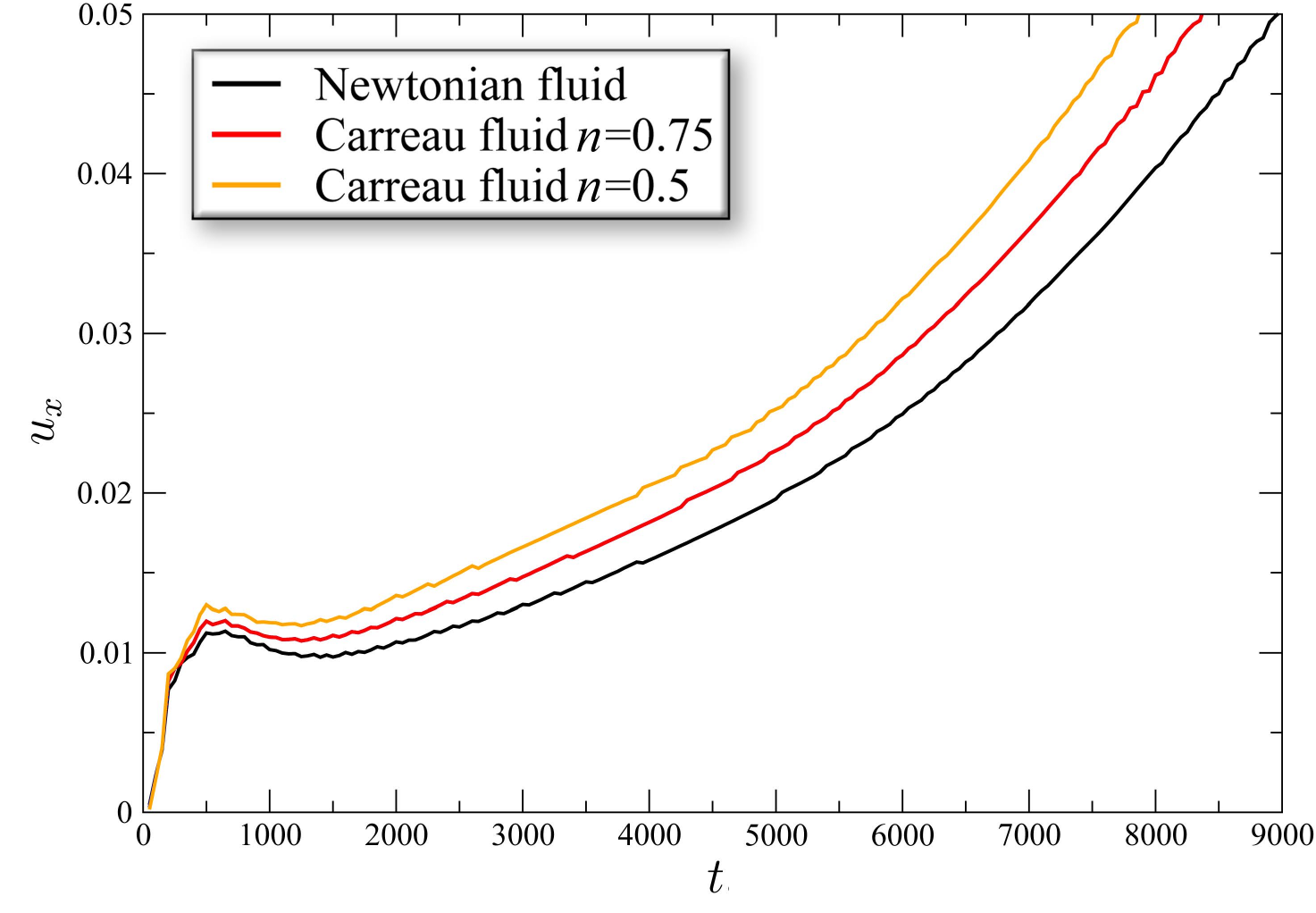}

\par\end{centering}

\protect\caption{Velocity component $u_x$ registered at the extreme point (rightest point) of the drop surface versus time
for all the three cases under investigation.}

\label{fig:07}
\end{figure}

We now report in Fig. \ref{fig:06} several snapshots taken over the time evolution of the system labelled $case \; b$.
In all the cases, we observe the formation of a liquid charged jet, which is ejected from the nozzle.
Further, we show in Fig. \ref{fig:07} the velocity component $u_x$ measured at the extreme (rightest) point of the drop surface versus time $t$.
Here, for all cases under investigation, we note that the velocity trend show the presence of a quasi-stationary point 
(see Fig. \ref{fig:06} panel a), where the viscous forces
balances the external electric force in agreement with previous theoretical investigations \cite{lauricella2015nonlinear,lauricella2015different,lauricella2017effects,reneker2000bending}.
In particular, such
quasi-stationary point is dependent on the rheology via the viscous stress,
since a different viscosity alters the balance point with the electric forces, 
providing a time shift of such regime.
Indeed, the simulation $case \; c$ reaches the quasi-stationary condition in a shorter time (as shown in Fig. \ref{fig:07}),
since the viscous stress is weaker in sustaining the expanding electrostatic pressure in the liquid phase.

\begin{figure}
\begin{centering}
\includegraphics[width=0.95\columnwidth]{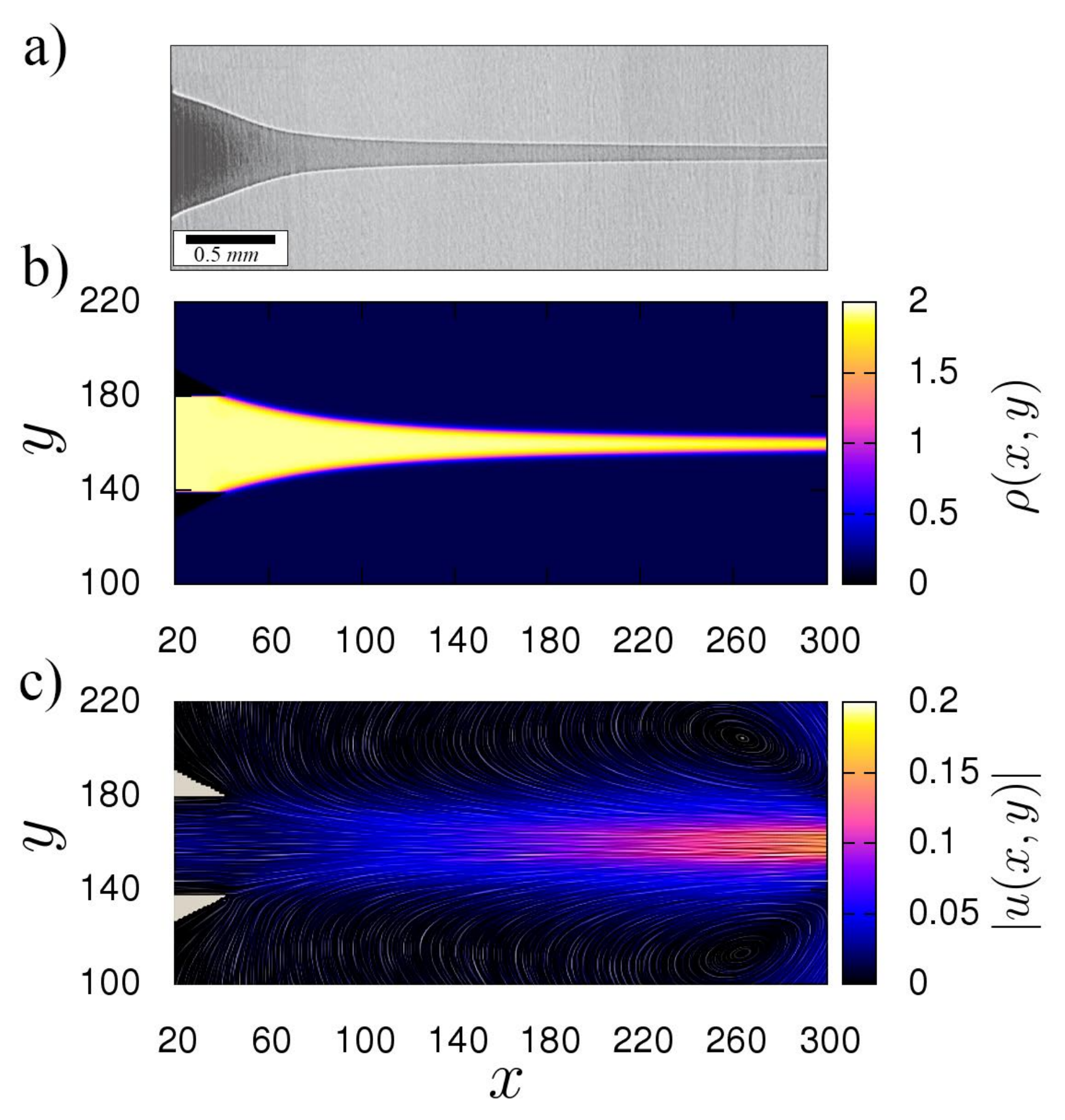}

\par\end{centering}

\protect\caption{In panel (a), a rectilinear section of a jet in an electrospinning experiment of a solution of 5 wt\% polyethylene oxide in water. Figure adapted with permission from Ref. \cite{greenfeld2011polymer}.
Copyrighted by the American Physical Society.
In panel (b), a snapshot of the fluid density $\rho$ in the stationary regime
after the jet has touched the right side of the simulation box
in the $case \; b$. In panel (c), the corresponding velocity field magnitude $|u(x,y)|$ and the LIC representation of the velocity field.}

\label{fig:08}
\end{figure}

After the jet touches the collector, the jet shape
fluctuates around a mean profile, providing a stationary regime.
In particular, at this stage the jet shows a hyperbolic profile (see Fig. \ref{fig:08} panel b) which appears to be in qualitative agreement
with the characteristic shape of the jet experimentally observed close to the injecting nozzle by the Rafailovich and Zussman groups
\cite{greenfeld2011polymer} (see Fig. \ref{fig:08} panel a)
and in consistency with previous theoretical results on the jet conical shape
\cite{reznik2010capillary,feng2003stretching,higuera2003flow,hohman2001electrospinning2,hartman1999electrohydrodynamic,cherney1999structure}.

\begin{figure}
\begin{centering}
\includegraphics[width=0.95\columnwidth]{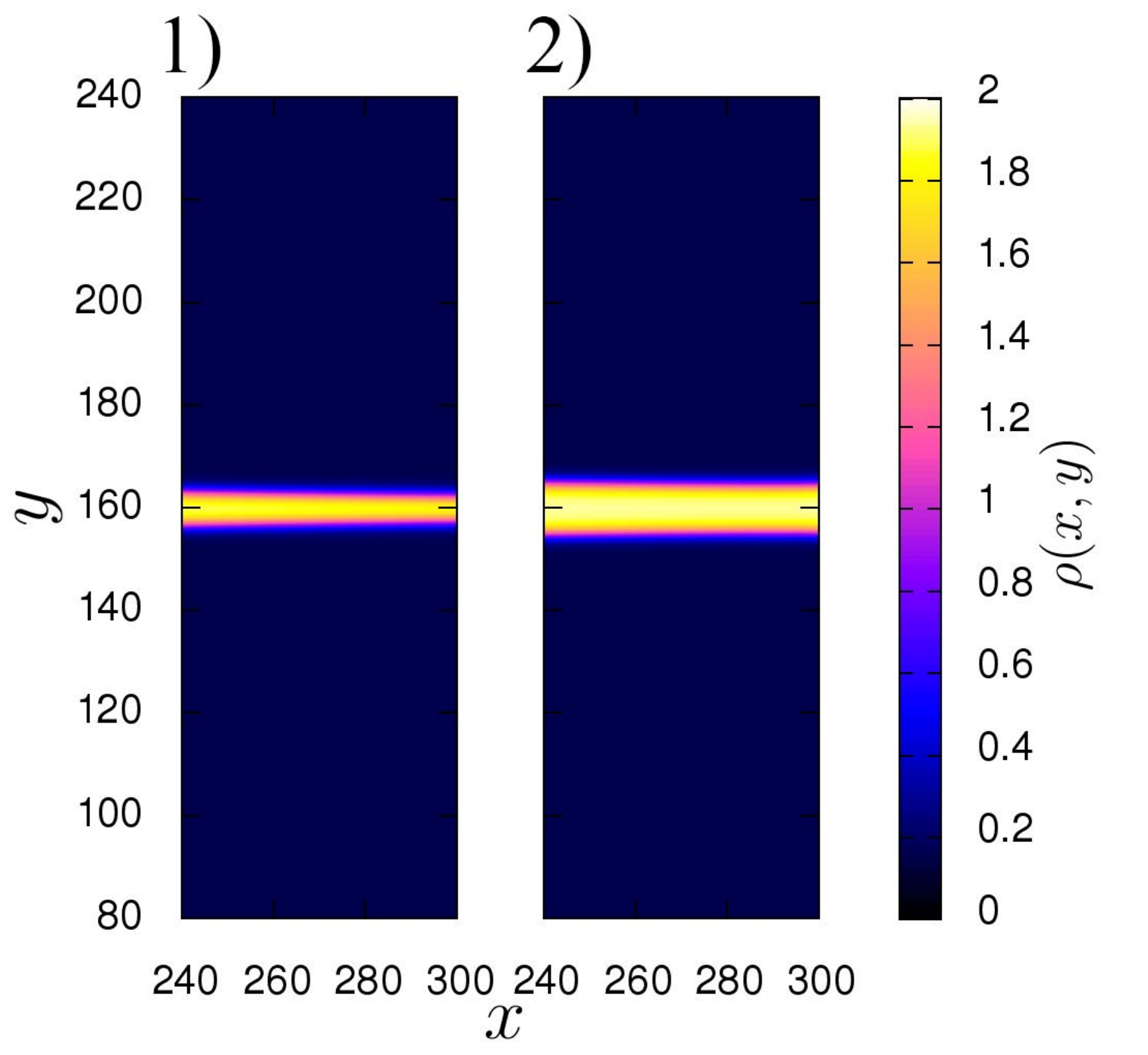}

\par\end{centering}

\protect\caption{Two sections of the jet profile $\rho(x,y)$ at timestep 10000 for the case $b$ (left panel 1), and for a simulation
without the local charge density and the Carreau model terms included in the ELBM (right panel 2). }

\label{fig:09}
\end{figure}

In Fig. \ref{fig:09}, the effects of the local charge density and the Carreau model terms in the present ELBM are investigated.
Here, we note that the jet diameter is quite sensitive to the inclusion of such effects. In particular, we observe a larger jet
diameter (see right panel of Fig. \ref{fig:09}), whenever these terms are not included in the model, showing a model
failure in minimizing the jet width.

To better compare our results with experimental data from the literature, we report the Ohnesorge number,
which describes the inertial, elastic and the capillary force balance.
Similarly, we exploit the Deborah number, which relates the elastic stress relaxation time to the Rayleigh timescale for the inertial-capillary breakup
of an inviscid jet.

In the context of the electrospinning process \cite{arinstein2017electrospun},
the Ohnesorge and Deborah numbers are
\begin{eqnarray}
Oh=\nu_0 \sqrt{\frac{\rho }{\sigma a_0}},  \nonumber  \\
De=\lambda \sqrt{\frac{\sigma}{\rho a^3_0}},
\label{eq:oh}
\end{eqnarray}
where $\rho$ is the mass density of the jet, $\sigma$ is the surface 
tension, 
$\nu_0$ is the zero shear kinematic viscosity, 
$\lambda$ is the relaxation time (see Subsec. \ref{subsec2}), 
and $a_0$ is the initial radius of the jet.

In our simulation, we estimate $Oh \approx 0.22$, and $De \approx 1.93$ for the
Ohnesorge and Deborah number, respectively, while 
these two dimensionless numbers are in the value ranges $Oh \approx 0.1 \-- 5.0$ and  $De \approx 0.1 \-- 30$
for a typical electrospinning scenario \cite{thompson2007effects,christanti2001surface}.

In order to characterize the stationary regime, we report the mean values of several observables measured
at the inlet. We take as a reference point the Cartesian coordinate $(x=1,z=160)$, which corresponds to the centre
of the nozzle diameter. Here, we measure the velocity along $\vec{x}$ equal to 
$2.8 \cdot 10^{-2}$ , $5.3  \cdot 10^{-2}$, and $6.1  \cdot 10^{-2}$ in lattice units, for the cases labeled $a$, $b$, and $c$, respectively.
Further, we observe in all the three cases almost the same values in the electric field along $\vec{x}$ equal to 
$5  \cdot 10^{-2}$ in lattice units, so we explain the velocity trend as a consequence of 
the lower viscosity depending on the different rheology in the three cases.

\begin{figure}
\begin{centering}
\includegraphics[width=0.95\columnwidth]{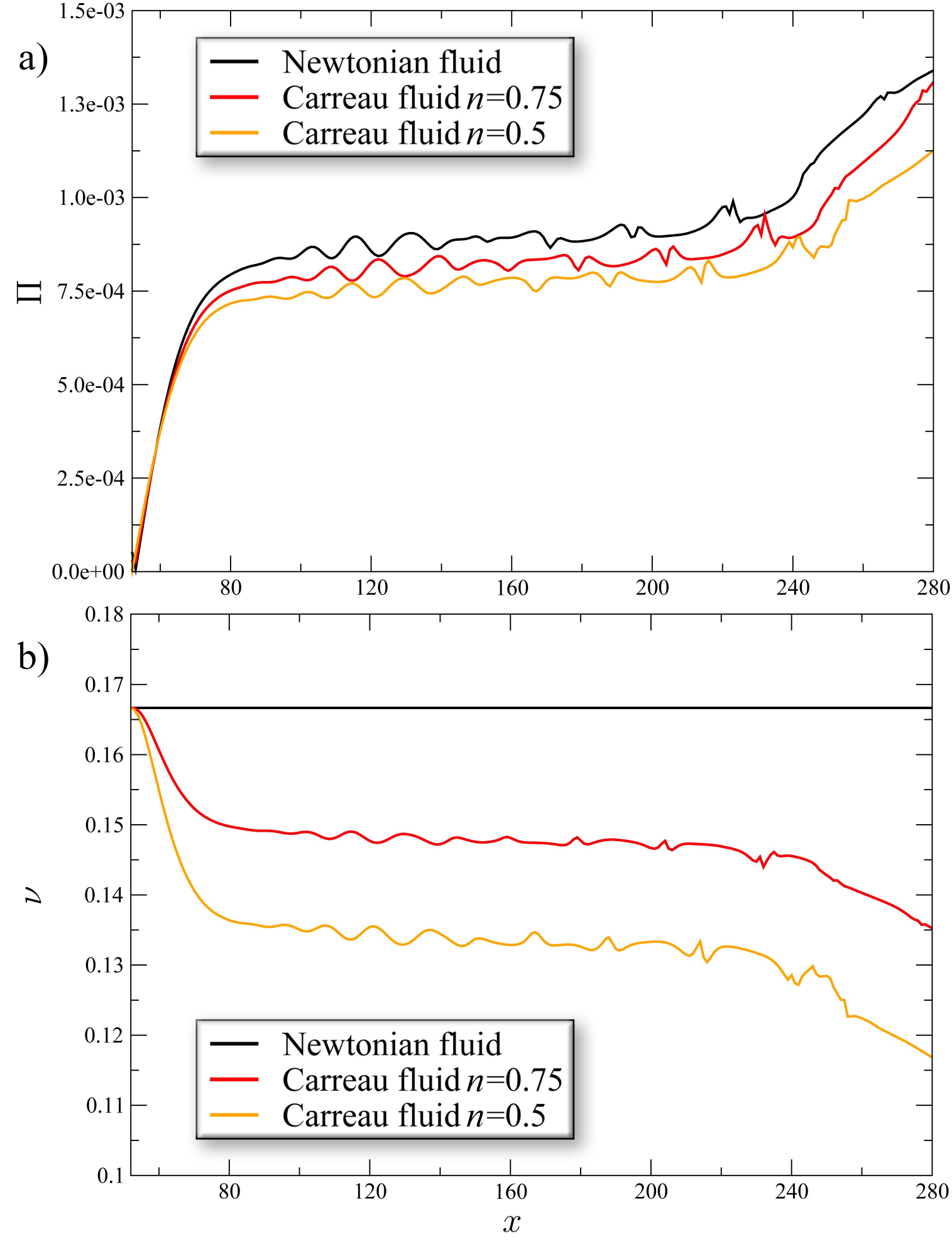}

\par\end{centering}

\protect\caption{In panel (a), the mean value of the stress $\Pi$ computed as matrix 2-norm $<||\mathbf{\Pi}||_2>_t$ of the stress tensor measured along the central axis of the jet $y=160$ averaged over a time interval of $15000$ steps.
In panel (b), the mean value of the kinematic viscosity $\nu$ computed with Eq. \ref{eq:carreau} along the central axis of the jet $y=160$, and averaged over a time interval of $15000$ steps.}

\label{fig:10}
\end{figure}

In particular, we observe a drag effect that is not depending on the local kinematic viscosity value at the inlet 
($\nu_{inlet} \approx 1/6$  in all the three cases).
Instead, this is due to the lower viscous force acting along the jet path outside the nozzle (as shown in panel b of Fig. \ref{fig:10}).

In order to characterize the stationary regime, we report in panel c of Fig. \ref{fig:08} the magnitude of the velocity field, and the line integral convolution (LIC) 
visualization technique \cite{cabral1993imaging}, highlighting the fine details of the flow field.
As expected, we observe a higher value of $u(x,z)$ along the jet towards the collector.
In particular, we analyze the profile of the velocity in a jet cross section along 
what is generally observed in the experimental process.

We investigate the effect of pseudoplastic rheology on the stress tensor $\mathbf{\Pi}$.
In panel a of Fig. \ref{fig:10} we report the mean value of the stress $\Pi$ measured along the
central axis of the jet $y=160$, and averaged over a time interval of $15000$ steps
in the stationary regime for the three cases under investigation. 

For all cases shown in Fig. \ref{fig:10}, we observe the presence of a drift in the $\mathbf{\Pi}$ profile starting from $x=240$.
This is mainly due to the larger magnitude of the external electric force,
which is originated by an increase in the surface-to-volume ratio. 
Since in a leaky dielectric the charge density
lies mainly on the surface, such increase in the surface-to-volume ratio provides a growth of the charge-to-mass ratio.
As a consequence, the jet undergoes a further stretching.

In Fig. \ref{fig:10}, we also note a decreasing trend of the stress $\mathbf{\Pi}$ by increasing the pseudoplasticity of the fluid (decreasing the flow index $n$).
Nonetheless, we observe a small shift in the stress magnitude. 
This is essentially due to the low value of relaxation time $\lambda$ in the Carreau model adopted in our simulation,
which provides a small decrease in the kinematic viscosity $\nu$. In order to clarify this point,
we report in panel b of Fig. \ref{fig:10} the mean value of the kinematic viscosity $<\nu>$ again assessed along the
central axis of the jet $y=160$, and averaged over the stationary regime time, where we observe
a small decrease of $<\nu>$ along the stretching direction as a function of the pseudoplastic
behavior in the fluid. These results look promising, and we plan to investigate systematically the effect of the rheological parameters on the jet dynamics in a future work.

\section{Summary}
Summarizing, we developed a Shan-Chen model for charged leaky dielectric fluids mainly aimed
in modelling the electrospinning process. The curvature effects on the charged surface were included
in our theoretical treatment, and we generalized the model to non-Newtonian flows
in order to account for the peculiar rheological behaviour.
Different scenarios were investigated to test the model. We initially investigated the
effect of strong electric fields on the droplet shape evolution. We also probed the jet formation under electrospinning-like conditions,
obtaining a good agreement with both experimental results and previous theoretical works present in literature.
At first glance, the pseudoplastic behaviour alters the jet dynamics, although a more systematic investigation requires an extensive test of the rheological parameters.
Work along these lines is currently underway.

The preliminary applications of the presented ELBM look promising, 
although more systematic numerical investigations, as well
as theoretical analysis, need to be undertaken. 
Nonetheless, the actual effort can be regarded as a significant forward step to
extend the applicability of the ELBM to the context of electrospinning systems,
providing a useful computational tool in the completion of the others presently available in the literature.

\section*{Acknowledgments}
The research leading to these results has received funding from the
European Research Council under the European Union's Seventh Framework
Programme (FP/2007-2013)/ERC Grant Agreement n. 306357 (\textquotedbl{}NANO-JETS\textquotedbl{}),
and under the European
Union's Horizon 2020 Framework Programme (No. FP/2014-
2020)/ERC Grant Agreement No. 739964  (\textquotedbl{}COPMAT\textquotedbl{}).

\revappendix*
\section{Units Conversion Table} 
\label{app:table}

\begin{table}[!hbtp]
\centering 
\scalebox{0.8}{
\begin{tabular}{c c c c}
\hline\hline 
Symbol & Definition & \thead{2-d lattice \\ units} & \thead{3-d cgs \\ units}  \\
\hline
$\rho$ & Mass density & $1\;\Delta m\,\Delta l^{-2}$  & $0.5 \; \text{g}\,\text{cm}^{-3} $\\
$u$ & Velocity  & $1\;\Delta l \, \Delta t^{-1}$ &  $1200 \; \text{cm} \,\text{s}^{-1}$\\
$\nu$ & Kinematic viscosity  & $1\;\Delta l^{2}\, \Delta t^{-1}$ & $1.2 \; \text{cm}^{2} \, \text{s}^{-1}$  \\
$q$ & Charge & $1\;\Delta l^{3/2} \, \Delta m^{1/2} \, \Delta t^{-1}$ &  $8.5 \cdot 10^{-4} \, \text{statC}$\\
$E$ & Electric field & $1\;\Delta l^{-1/2}\, \Delta m^{1/2}\, \Delta t^{-1}$ & $848.5 \; \text{statV} \, \text{cm}^{-1} $\\
$\phi$ & Electric potential & $1\;\Delta l^{1/2}\, \Delta m^{1/2}\, \Delta t^{-1}$ & $8.485 \cdot 10^{-1} \; \text{statV}$ \\
$\varepsilon$ & Relative permittivity & $-$ & $-$ \\
$a_0$ & Initial jet radius & $1\;\Delta l$ & $10^{-3}  \text{cm}$ \\
$K$ & Mean curvature & $1\; \Delta l^{-1}$ & $1000 \; \text{cm}^{-1} $ \\
$\dot{\gamma}$ & Shear rate & $1\; \Delta t^{-1}$ & $1.2 \cdot 10^6 \, \text{s}^{-1}$ \\
$\Pi$ & Stress & $1\; \Delta m \, \Delta t^{-2}$ &  $7.2 \cdot 10^5 \, \text{g}\,\text{cm}^{-1} \, \text{s}^{-2}$\\
$\lambda$ & Relaxation time & $1\; \Delta t$ & $8.3 \cdot 10^{-7} \; \text{s}$ \\
$n$ & Index flow& $-$ & $-$\\
$p$ & Pressure & $1\; \Delta m \, \Delta t^{-2}$ &  $7.2 \cdot 10^5 \, \text{g}\,\text{cm}^{-1} \, \text{s}^{-2}$ \\
$\sigma$ & Surface tension & $1\; \Delta l \, \Delta m \, \Delta t^{-2}$ & $720 \; \text{g} \, \text{s}^{-2}$\\
\hline
\end{tabular}}
\label{tab:unit}
\caption{Symbols employed in order of appearance, their definitions, relative dimensions in 2-d lattice units, and conversion values in 3-d Gaussian centimetre-gram-second (cgs) system of units. 
The conversion to 3-d dimension units is obtained by multiplying the 2-d square area $\Delta l^2$  for an extra node $\Delta l $ along the z-axis. 
Note that $\Delta m$ denotes the mass unit equal to $5 \cdot 10^{-10}$ g (water density), $\Delta l$ the length unit equal to $10^{-3}  \text{cm}$, 
and time step $\Delta t$ is the time unit equal to $8.3 \cdot 10^{-7}$ s.  The last number was tuned in order to set the kinematic viscosity  $\nu=1/6 \; \Delta l^{2} \, \Delta t^{-1} = 0.2 \; \text{cm}^2 \text{s}^{-1}$ in similarity with the viscosity
value adopted in the Lagrangian model of Ref. \cite{lauricella2015jetspin}.}
\end{table}


\begin{thebibliography}{66}%
\makeatletter
\providecommand \@ifxundefined [1]{%
 \@ifx{#1\undefined}
}%
\providecommand \@ifnum [1]{%
 \ifnum #1\expandafter \@firstoftwo
 \else \expandafter \@secondoftwo
 \fi
}%
\providecommand \@ifx [1]{%
 \ifx #1\expandafter \@firstoftwo
 \else \expandafter \@secondoftwo
 \fi
}%
\providecommand \natexlab [1]{#1}%
\providecommand \enquote  [1]{``#1''}%
\providecommand \bibnamefont  [1]{#1}%
\providecommand \bibfnamefont [1]{#1}%
\providecommand \citenamefont [1]{#1}%
\providecommand \href@noop [0]{\@secondoftwo}%
\providecommand \href [0]{\begingroup \@sanitize@url \@href}%
\providecommand \@href[1]{\@@startlink{#1}\@@href}%
\providecommand \@@href[1]{\endgroup#1\@@endlink}%
\providecommand \@sanitize@url [0]{\catcode `\\12\catcode `\$12\catcode
  `\&12\catcode `\#12\catcode `\^12\catcode `\_12\catcode `\%12\relax}%
\providecommand \@@startlink[1]{}%
\providecommand \@@endlink[0]{}%
\providecommand \url  [0]{\begingroup\@sanitize@url \@url }%
\providecommand \@url [1]{\endgroup\@href {#1}{\urlprefix }}%
\providecommand \urlprefix  [0]{URL }%
\providecommand \Eprint [0]{\href }%
\providecommand \doibase [0]{http://dx.doi.org/}%
\providecommand \selectlanguage [0]{\@gobble}%
\providecommand \bibinfo  [0]{\@secondoftwo}%
\providecommand \bibfield  [0]{\@secondoftwo}%
\providecommand \translation [1]{[#1]}%
\providecommand \BibitemOpen [0]{}%
\providecommand \bibitemStop [0]{}%
\providecommand \bibitemNoStop [0]{.\EOS\space}%
\providecommand \EOS [0]{\spacefactor3000\relax}%
\providecommand \BibitemShut  [1]{\csname bibitem#1\endcsname}%
\let\auto@bib@innerbib\@empty
\bibitem [{\citenamefont {Yarin}\ \emph {et~al.}(2014)\citenamefont {Yarin},
  \citenamefont {Pourdeyhimi},\ and\ \citenamefont
  {Ramakrishna}}]{yarin2014fundamentals}%
  \BibitemOpen
  \bibfield  {author} {\bibinfo {author} {\bibfnamefont {A.~L.}\ \bibnamefont
  {Yarin}}, \bibinfo {author} {\bibfnamefont {B.}~\bibnamefont {Pourdeyhimi}},
  \ and\ \bibinfo {author} {\bibfnamefont {S.}~\bibnamefont {Ramakrishna}},\
  }\href@noop {} {\emph {\bibinfo {title} {Fundamentals and applications of
  micro-and nanofibers}}}\ (\bibinfo  {publisher} {Cambridge University
  Press},\ \bibinfo {year} {2014})\BibitemShut {NoStop}%
\bibitem [{\citenamefont {Wendorff}\ \emph {et~al.}(2012)\citenamefont
  {Wendorff}, \citenamefont {Agarwal},\ and\ \citenamefont
  {Greiner}}]{wendorff2012electrospinning}%
  \BibitemOpen
  \bibfield  {author} {\bibinfo {author} {\bibfnamefont {J.~H.}\ \bibnamefont
  {Wendorff}}, \bibinfo {author} {\bibfnamefont {S.}~\bibnamefont {Agarwal}}, \
  and\ \bibinfo {author} {\bibfnamefont {A.}~\bibnamefont {Greiner}},\
  }\href@noop {} {\emph {\bibinfo {title} {Electrospinning: materials,
  processing, and applications}}}\ (\bibinfo  {publisher} {John Wiley \&
  Sons},\ \bibinfo {year} {2012})\BibitemShut {NoStop}%
\bibitem [{\citenamefont {Pisignano}(2013)}]{pisignano2013polymer}%
  \BibitemOpen
  \bibfield  {author} {\bibinfo {author} {\bibfnamefont {D.}~\bibnamefont
  {Pisignano}},\ }\href@noop {} {\emph {\bibinfo {title} {Polymer Nanofibers:
  Building Blocks for Nanotechnology}}},\ Vol.~\bibinfo {volume} {29}\
  (\bibinfo  {publisher} {Royal Society of Chemistry},\ \bibinfo {year}
  {2013})\BibitemShut {NoStop}%
\bibitem [{\citenamefont {Frenot}\ and\ \citenamefont
  {Chronakis}(2003)}]{frenot2003polymer}%
  \BibitemOpen
  \bibfield  {author} {\bibinfo {author} {\bibfnamefont {A.}~\bibnamefont
  {Frenot}}\ and\ \bibinfo {author} {\bibfnamefont {I.~S.}\ \bibnamefont
  {Chronakis}},\ }\href@noop {} {\bibfield  {journal} {\bibinfo  {journal}
  {Current opinion in colloid \& interface science}\ }\textbf {\bibinfo
  {volume} {8}},\ \bibinfo {pages} {64} (\bibinfo {year} {2003})}\BibitemShut
  {NoStop}%
\bibitem [{\citenamefont {Huang}\ \emph {et~al.}(2003)\citenamefont {Huang},
  \citenamefont {Zhang}, \citenamefont {Kotaki},\ and\ \citenamefont
  {Ramakrishna}}]{huang2003review}%
  \BibitemOpen
  \bibfield  {author} {\bibinfo {author} {\bibfnamefont {Z.-M.}\ \bibnamefont
  {Huang}}, \bibinfo {author} {\bibfnamefont {Y.-Z.}\ \bibnamefont {Zhang}},
  \bibinfo {author} {\bibfnamefont {M.}~\bibnamefont {Kotaki}}, \ and\ \bibinfo
  {author} {\bibfnamefont {S.}~\bibnamefont {Ramakrishna}},\ }\href@noop {}
  {\bibfield  {journal} {\bibinfo  {journal} {Composites science and
  technology}\ }\textbf {\bibinfo {volume} {63}},\ \bibinfo {pages} {2223}
  (\bibinfo {year} {2003})}\BibitemShut {NoStop}%
\bibitem [{\citenamefont {Li}\ \emph {et~al.}(2004)\citenamefont {Li},
  \citenamefont {Wang},\ and\ \citenamefont {Xia}}]{li2004electrospinning}%
  \BibitemOpen
  \bibfield  {author} {\bibinfo {author} {\bibfnamefont {D.}~\bibnamefont
  {Li}}, \bibinfo {author} {\bibfnamefont {Y.}~\bibnamefont {Wang}}, \ and\
  \bibinfo {author} {\bibfnamefont {Y.}~\bibnamefont {Xia}},\ }\href@noop {}
  {\bibfield  {journal} {\bibinfo  {journal} {Advanced Materials}\ }\textbf
  {\bibinfo {volume} {16}},\ \bibinfo {pages} {361} (\bibinfo {year}
  {2004})}\BibitemShut {NoStop}%
\bibitem [{\citenamefont {Greiner}\ and\ \citenamefont
  {Wendorff}(2007)}]{greiner2007electrospinning}%
  \BibitemOpen
  \bibfield  {author} {\bibinfo {author} {\bibfnamefont {A.}~\bibnamefont
  {Greiner}}\ and\ \bibinfo {author} {\bibfnamefont {J.~H.}\ \bibnamefont
  {Wendorff}},\ }\href@noop {} {\bibfield  {journal} {\bibinfo  {journal}
  {Angewandte Chemie International Edition}\ }\textbf {\bibinfo {volume}
  {46}},\ \bibinfo {pages} {5670} (\bibinfo {year} {2007})}\BibitemShut
  {NoStop}%
\bibitem [{\citenamefont {Carroll}\ \emph {et~al.}(2008)\citenamefont
  {Carroll}, \citenamefont {Zhmayev}, \citenamefont {Kalra}, \citenamefont
  {Joo} \emph {et~al.}}]{carroll2008nanofibers}%
  \BibitemOpen
  \bibfield  {author} {\bibinfo {author} {\bibfnamefont {C.~P.}\ \bibnamefont
  {Carroll}}, \bibinfo {author} {\bibfnamefont {E.}~\bibnamefont {Zhmayev}},
  \bibinfo {author} {\bibfnamefont {V.}~\bibnamefont {Kalra}}, \bibinfo
  {author} {\bibfnamefont {Y.~L.}\ \bibnamefont {Joo}},  \emph {et~al.},\
  }\href@noop {} {\bibfield  {journal} {\bibinfo  {journal} {Korea-Australia
  Rheology Journal}\ }\textbf {\bibinfo {volume} {20}},\ \bibinfo {pages} {153}
  (\bibinfo {year} {2008})}\BibitemShut {NoStop}%
\bibitem [{\citenamefont {Persano}\ \emph {et~al.}(2013)\citenamefont
  {Persano}, \citenamefont {Camposeo}, \citenamefont {Tekmen},\ and\
  \citenamefont {Pisignano}}]{persano2013industrial}%
  \BibitemOpen
  \bibfield  {author} {\bibinfo {author} {\bibfnamefont {L.}~\bibnamefont
  {Persano}}, \bibinfo {author} {\bibfnamefont {A.}~\bibnamefont {Camposeo}},
  \bibinfo {author} {\bibfnamefont {C.}~\bibnamefont {Tekmen}}, \ and\ \bibinfo
  {author} {\bibfnamefont {D.}~\bibnamefont {Pisignano}},\ }\href@noop {}
  {\bibfield  {journal} {\bibinfo  {journal} {Macromolecular Materials and
  Engineering}\ }\textbf {\bibinfo {volume} {298}},\ \bibinfo {pages} {504}
  (\bibinfo {year} {2013})}\BibitemShut {NoStop}%
\bibitem [{\citenamefont {Reneker}\ \emph {et~al.}(2000)\citenamefont
  {Reneker}, \citenamefont {Yarin}, \citenamefont {Fong},\ and\ \citenamefont
  {Koombhongse}}]{reneker2000bending}%
  \BibitemOpen
  \bibfield  {author} {\bibinfo {author} {\bibfnamefont {D.~H.}\ \bibnamefont
  {Reneker}}, \bibinfo {author} {\bibfnamefont {A.~L.}\ \bibnamefont {Yarin}},
  \bibinfo {author} {\bibfnamefont {H.}~\bibnamefont {Fong}}, \ and\ \bibinfo
  {author} {\bibfnamefont {S.}~\bibnamefont {Koombhongse}},\ }\href@noop {}
  {\bibfield  {journal} {\bibinfo  {journal} {Journal of Applied physics}\
  }\textbf {\bibinfo {volume} {87}},\ \bibinfo {pages} {4531} (\bibinfo {year}
  {2000})}\BibitemShut {NoStop}%
\bibitem [{\citenamefont {Feng}(2003)}]{feng2003stretching}%
  \BibitemOpen
  \bibfield  {author} {\bibinfo {author} {\bibfnamefont {J.}~\bibnamefont
  {Feng}},\ }\href@noop {} {\bibfield  {journal} {\bibinfo  {journal} {Journal
  of Non-Newtonian Fluid Mechanics}\ }\textbf {\bibinfo {volume} {116}},\
  \bibinfo {pages} {55} (\bibinfo {year} {2003})}\BibitemShut {NoStop}%
\bibitem [{\citenamefont {Lauricella}\ \emph
  {et~al.}(2015{\natexlab{a}})\citenamefont {Lauricella}, \citenamefont
  {Pontrelli}, \citenamefont {Coluzza}, \citenamefont {Pisignano},\ and\
  \citenamefont {Succi}}]{lauricella2015jetspin}%
  \BibitemOpen
  \bibfield  {author} {\bibinfo {author} {\bibfnamefont {M.}~\bibnamefont
  {Lauricella}}, \bibinfo {author} {\bibfnamefont {G.}~\bibnamefont
  {Pontrelli}}, \bibinfo {author} {\bibfnamefont {I.}~\bibnamefont {Coluzza}},
  \bibinfo {author} {\bibfnamefont {D.}~\bibnamefont {Pisignano}}, \ and\
  \bibinfo {author} {\bibfnamefont {S.}~\bibnamefont {Succi}},\ }\href@noop {}
  {\bibfield  {journal} {\bibinfo  {journal} {Computer Physics Communications}\
  }\textbf {\bibinfo {volume} {197}},\ \bibinfo {pages} {227} (\bibinfo {year}
  {2015}{\natexlab{a}})}\BibitemShut {NoStop}%
\bibitem [{\citenamefont {Lauricella}\ \emph
  {et~al.}(2016{\natexlab{a}})\citenamefont {Lauricella}, \citenamefont
  {Pontrelli}, \citenamefont {Pisignano},\ and\ \citenamefont
  {Succi}}]{lauricella2016dynamic}%
  \BibitemOpen
  \bibfield  {author} {\bibinfo {author} {\bibfnamefont {M.}~\bibnamefont
  {Lauricella}}, \bibinfo {author} {\bibfnamefont {G.}~\bibnamefont
  {Pontrelli}}, \bibinfo {author} {\bibfnamefont {D.}~\bibnamefont
  {Pisignano}}, \ and\ \bibinfo {author} {\bibfnamefont {S.}~\bibnamefont
  {Succi}},\ }\href@noop {} {\bibfield  {journal} {\bibinfo  {journal} {Journal
  of Computational Science}\ }\textbf {\bibinfo {volume} {17}},\ \bibinfo
  {pages} {325} (\bibinfo {year} {2016}{\natexlab{a}})}\BibitemShut {NoStop}%
\bibitem [{\citenamefont {Lauricella}\ \emph
  {et~al.}(2016{\natexlab{b}})\citenamefont {Lauricella}, \citenamefont
  {Pisignano},\ and\ \citenamefont {Succi}}]{lauricella2016three}%
  \BibitemOpen
  \bibfield  {author} {\bibinfo {author} {\bibfnamefont {M.}~\bibnamefont
  {Lauricella}}, \bibinfo {author} {\bibfnamefont {D.}~\bibnamefont
  {Pisignano}}, \ and\ \bibinfo {author} {\bibfnamefont {S.}~\bibnamefont
  {Succi}},\ }\href@noop {} {\bibfield  {journal} {\bibinfo  {journal} {The
  journal of physical chemistry. A}\ }\textbf {\bibinfo {volume} {120}},\
  \bibinfo {pages} {4884} (\bibinfo {year} {2016}{\natexlab{b}})}\BibitemShut
  {NoStop}%
\bibitem [{\citenamefont {Ganan-Calvo}(1997)}]{ganan1997theory}%
  \BibitemOpen
  \bibfield  {author} {\bibinfo {author} {\bibfnamefont {A.~M.}\ \bibnamefont
  {Ganan-Calvo}},\ }\href@noop {} {\bibfield  {journal} {\bibinfo  {journal}
  {Journal of Fluid Mechanics}\ }\textbf {\bibinfo {volume} {335}},\ \bibinfo
  {pages} {165} (\bibinfo {year} {1997})}\BibitemShut {NoStop}%
\bibitem [{\citenamefont {Hohman}\ \emph
  {et~al.}(2001{\natexlab{a}})\citenamefont {Hohman}, \citenamefont {Shin},
  \citenamefont {Rutledge},\ and\ \citenamefont
  {Brenner}}]{hohman2001electrospinning}%
  \BibitemOpen
  \bibfield  {author} {\bibinfo {author} {\bibfnamefont {M.~M.}\ \bibnamefont
  {Hohman}}, \bibinfo {author} {\bibfnamefont {M.}~\bibnamefont {Shin}},
  \bibinfo {author} {\bibfnamefont {G.}~\bibnamefont {Rutledge}}, \ and\
  \bibinfo {author} {\bibfnamefont {M.~P.}\ \bibnamefont {Brenner}},\
  }\href@noop {} {\bibfield  {journal} {\bibinfo  {journal} {Physics of
  fluids}\ }\textbf {\bibinfo {volume} {13}},\ \bibinfo {pages} {2201}
  (\bibinfo {year} {2001}{\natexlab{a}})}\BibitemShut {NoStop}%
\bibitem [{\citenamefont {Yarin}\ \emph
  {et~al.}(2001{\natexlab{a}})\citenamefont {Yarin}, \citenamefont
  {Koombhongse},\ and\ \citenamefont {Reneker}}]{yarin2001bending}%
  \BibitemOpen
  \bibfield  {author} {\bibinfo {author} {\bibfnamefont {A.~L.}\ \bibnamefont
  {Yarin}}, \bibinfo {author} {\bibfnamefont {S.}~\bibnamefont {Koombhongse}},
  \ and\ \bibinfo {author} {\bibfnamefont {D.~H.}\ \bibnamefont {Reneker}},\
  }\href@noop {} {\bibfield  {journal} {\bibinfo  {journal} {Journal of applied
  physics}\ }\textbf {\bibinfo {volume} {89}},\ \bibinfo {pages} {3018}
  (\bibinfo {year} {2001}{\natexlab{a}})}\BibitemShut {NoStop}%
\bibitem [{\citenamefont {Li}\ and\ \citenamefont
  {Kwok}(2003)}]{li2003discrete}%
  \BibitemOpen
  \bibfield  {author} {\bibinfo {author} {\bibfnamefont {B.}~\bibnamefont
  {Li}}\ and\ \bibinfo {author} {\bibfnamefont {D.~Y.}\ \bibnamefont {Kwok}},\
  }\href@noop {} {\bibfield  {journal} {\bibinfo  {journal} {Physical review
  letters}\ }\textbf {\bibinfo {volume} {90}},\ \bibinfo {pages} {124502}
  (\bibinfo {year} {2003})}\BibitemShut {NoStop}%
\bibitem [{\citenamefont {Kupershtokh}\ and\ \citenamefont
  {Medvedev}(2006)}]{kupershtokh2006lattice}%
  \BibitemOpen
  \bibfield  {author} {\bibinfo {author} {\bibfnamefont {A.}~\bibnamefont
  {Kupershtokh}}\ and\ \bibinfo {author} {\bibfnamefont {D.}~\bibnamefont
  {Medvedev}},\ }\href@noop {} {\bibfield  {journal} {\bibinfo  {journal}
  {Journal of electrostatics}\ }\textbf {\bibinfo {volume} {64}},\ \bibinfo
  {pages} {581} (\bibinfo {year} {2006})}\BibitemShut {NoStop}%
\bibitem [{\citenamefont {Huang}\ \emph {et~al.}(2007)\citenamefont {Huang},
  \citenamefont {Li},\ and\ \citenamefont {Liu}}]{huang2007application}%
  \BibitemOpen
  \bibfield  {author} {\bibinfo {author} {\bibfnamefont {W.}~\bibnamefont
  {Huang}}, \bibinfo {author} {\bibfnamefont {Y.}~\bibnamefont {Li}}, \ and\
  \bibinfo {author} {\bibfnamefont {Q.}~\bibnamefont {Liu}},\ }\href@noop {}
  {\bibfield  {journal} {\bibinfo  {journal} {Chinese Science Bulletin}\
  }\textbf {\bibinfo {volume} {52}},\ \bibinfo {pages} {3319} (\bibinfo {year}
  {2007})}\BibitemShut {NoStop}%
\bibitem [{\citenamefont {Gong}\ \emph {et~al.}(2010)\citenamefont {Gong},
  \citenamefont {Cheng},\ and\ \citenamefont {Quan}}]{gong2010lattice}%
  \BibitemOpen
  \bibfield  {author} {\bibinfo {author} {\bibfnamefont {S.}~\bibnamefont
  {Gong}}, \bibinfo {author} {\bibfnamefont {P.}~\bibnamefont {Cheng}}, \ and\
  \bibinfo {author} {\bibfnamefont {X.}~\bibnamefont {Quan}},\ }\href@noop {}
  {\bibfield  {journal} {\bibinfo  {journal} {International Journal of Heat and
  Mass Transfer}\ }\textbf {\bibinfo {volume} {53}},\ \bibinfo {pages} {5863}
  (\bibinfo {year} {2010})}\BibitemShut {NoStop}%
\bibitem [{\citenamefont {Medvedev}(2010)}]{medvedev2010lattice}%
  \BibitemOpen
  \bibfield  {author} {\bibinfo {author} {\bibfnamefont {D.}~\bibnamefont
  {Medvedev}},\ }\href@noop {} {\bibfield  {journal} {\bibinfo  {journal}
  {Procedia Computer Science}\ }\textbf {\bibinfo {volume} {1}},\ \bibinfo
  {pages} {811} (\bibinfo {year} {2010})}\BibitemShut {NoStop}%
\bibitem [{\citenamefont {Bararnia}\ and\ \citenamefont
  {Ganji}(2013)}]{bararnia2013breakup}%
  \BibitemOpen
  \bibfield  {author} {\bibinfo {author} {\bibfnamefont {H.}~\bibnamefont
  {Bararnia}}\ and\ \bibinfo {author} {\bibfnamefont {D.}~\bibnamefont
  {Ganji}},\ }\href@noop {} {\bibfield  {journal} {\bibinfo  {journal}
  {Advanced Powder Technology}\ }\textbf {\bibinfo {volume} {24}},\ \bibinfo
  {pages} {992} (\bibinfo {year} {2013})}\BibitemShut {NoStop}%
\bibitem [{\citenamefont {Kupershtokh}(2014)}]{kupershtokh2014three}%
  \BibitemOpen
  \bibfield  {author} {\bibinfo {author} {\bibfnamefont {A.~L.}\ \bibnamefont
  {Kupershtokh}},\ }\href@noop {} {\bibfield  {journal} {\bibinfo  {journal}
  {Computers \& Mathematics with Applications}\ }\textbf {\bibinfo {volume}
  {67}},\ \bibinfo {pages} {340} (\bibinfo {year} {2014})}\BibitemShut
  {NoStop}%
\bibitem [{\citenamefont {Landau}\ \emph {et~al.}(2013)\citenamefont {Landau},
  \citenamefont {Bell}, \citenamefont {Kearsley}, \citenamefont {Pitaevskii},
  \citenamefont {Lifshitz},\ and\ \citenamefont
  {Sykes}}]{landau2013electrodynamics}%
  \BibitemOpen
  \bibfield  {author} {\bibinfo {author} {\bibfnamefont {L.~D.}\ \bibnamefont
  {Landau}}, \bibinfo {author} {\bibfnamefont {J.}~\bibnamefont {Bell}},
  \bibinfo {author} {\bibfnamefont {M.}~\bibnamefont {Kearsley}}, \bibinfo
  {author} {\bibfnamefont {L.}~\bibnamefont {Pitaevskii}}, \bibinfo {author}
  {\bibfnamefont {E.}~\bibnamefont {Lifshitz}}, \ and\ \bibinfo {author}
  {\bibfnamefont {J.}~\bibnamefont {Sykes}},\ }\href@noop {} {\emph {\bibinfo
  {title} {Electrodynamics of continuous media}}},\ Vol.~\bibinfo {volume} {8}\
  (\bibinfo  {publisher} {elsevier},\ \bibinfo {year} {2013})\BibitemShut
  {NoStop}%
\bibitem [{\citenamefont {Taylor}(1964)}]{taylor1964disintegration}%
  \BibitemOpen
  \bibfield  {author} {\bibinfo {author} {\bibfnamefont {G.}~\bibnamefont
  {Taylor}},\ }\bibfield  {booktitle} {\emph {\bibinfo {booktitle} {Proceedings
  of the Royal Society of London A: Mathematical, Physical and Engineering
  Sciences}},\ }\href@noop {} {\ \textbf {\bibinfo {volume} {280}},\ \bibinfo
  {pages} {383} (\bibinfo {year} {1964})}\BibitemShut {NoStop}%
\bibitem [{\citenamefont {Taylor}(1966)}]{taylor1966studies}%
  \BibitemOpen
  \bibfield  {author} {\bibinfo {author} {\bibfnamefont {G.}~\bibnamefont
  {Taylor}},\ }\bibfield  {booktitle} {\emph {\bibinfo {booktitle} {Proceedings
  of the Royal Society of London A: Mathematical, Physical and Engineering
  Sciences}},\ }\href@noop {} {\ \textbf {\bibinfo {volume} {291}},\ \bibinfo
  {pages} {159} (\bibinfo {year} {1966})}\BibitemShut {NoStop}%
\bibitem [{\citenamefont {Taylor}(1969)}]{taylor1969electrically}%
  \BibitemOpen
  \bibfield  {author} {\bibinfo {author} {\bibfnamefont {G.}~\bibnamefont
  {Taylor}},\ }\bibfield  {booktitle} {\emph {\bibinfo {booktitle} {Proceedings
  of the Royal Society of London A: Mathematical, Physical and Engineering
  Sciences}},\ }\href@noop {} {\ \textbf {\bibinfo {volume} {313}},\ \bibinfo
  {pages} {453} (\bibinfo {year} {1969})}\BibitemShut {NoStop}%
\bibitem [{\citenamefont {Saville}(1997)}]{saville1997electrohydrodynamics}%
  \BibitemOpen
  \bibfield  {author} {\bibinfo {author} {\bibfnamefont {D.}~\bibnamefont
  {Saville}},\ }\href@noop {} {\bibfield  {journal} {\bibinfo  {journal}
  {Annual review of fluid mechanics}\ }\textbf {\bibinfo {volume} {29}},\
  \bibinfo {pages} {27} (\bibinfo {year} {1997})}\BibitemShut {NoStop}%
\bibitem [{\citenamefont {Karlin}\ \emph {et~al.}(1999)\citenamefont {Karlin},
  \citenamefont {Ferrante},\ and\ \citenamefont
  {{\"O}ttinger}}]{karlin1999perfect}%
  \BibitemOpen
  \bibfield  {author} {\bibinfo {author} {\bibfnamefont {I.}~\bibnamefont
  {Karlin}}, \bibinfo {author} {\bibfnamefont {A.}~\bibnamefont {Ferrante}}, \
  and\ \bibinfo {author} {\bibfnamefont {H.~C.}\ \bibnamefont {{\"O}ttinger}},\
  }\href@noop {} {\bibfield  {journal} {\bibinfo  {journal} {EPL (Europhysics
  Letters)}\ }\textbf {\bibinfo {volume} {47}},\ \bibinfo {pages} {182}
  (\bibinfo {year} {1999})}\BibitemShut {NoStop}%
\bibitem [{\citenamefont {Hua}\ \emph {et~al.}(2008)\citenamefont {Hua},
  \citenamefont {Lim},\ and\ \citenamefont {Wang}}]{hua2008numerical}%
  \BibitemOpen
  \bibfield  {author} {\bibinfo {author} {\bibfnamefont {J.}~\bibnamefont
  {Hua}}, \bibinfo {author} {\bibfnamefont {L.~K.}\ \bibnamefont {Lim}}, \ and\
  \bibinfo {author} {\bibfnamefont {C.-H.}\ \bibnamefont {Wang}},\ }\href@noop
  {} {\bibfield  {journal} {\bibinfo  {journal} {Physics of Fluids}\ }\textbf
  {\bibinfo {volume} {20}},\ \bibinfo {pages} {113302} (\bibinfo {year}
  {2008})}\BibitemShut {NoStop}%
\bibitem [{\citenamefont {Li}\ \emph {et~al.}(2011)\citenamefont {Li},
  \citenamefont {Li}, \citenamefont {Huang},\ and\ \citenamefont
  {Lu}}]{li2011lattice}%
  \BibitemOpen
  \bibfield  {author} {\bibinfo {author} {\bibfnamefont {Z.-T.}\ \bibnamefont
  {Li}}, \bibinfo {author} {\bibfnamefont {G.-J.}\ \bibnamefont {Li}}, \bibinfo
  {author} {\bibfnamefont {H.-B.}\ \bibnamefont {Huang}}, \ and\ \bibinfo
  {author} {\bibfnamefont {X.-Y.}\ \bibnamefont {Lu}},\ }\href@noop {}
  {\bibfield  {journal} {\bibinfo  {journal} {International Journal of Modern
  Physics C}\ }\textbf {\bibinfo {volume} {22}},\ \bibinfo {pages} {729}
  (\bibinfo {year} {2011})}\BibitemShut {NoStop}%
\bibitem [{\citenamefont {Ansumali}\ and\ \citenamefont
  {Karlin}(2002)}]{ansumali2002single}%
  \BibitemOpen
  \bibfield  {author} {\bibinfo {author} {\bibfnamefont {S.}~\bibnamefont
  {Ansumali}}\ and\ \bibinfo {author} {\bibfnamefont {I.~V.}\ \bibnamefont
  {Karlin}},\ }\href@noop {} {\bibfield  {journal} {\bibinfo  {journal}
  {Physical Review E}\ }\textbf {\bibinfo {volume} {65}},\ \bibinfo {pages}
  {056312} (\bibinfo {year} {2002})}\BibitemShut {NoStop}%
\bibitem [{\citenamefont {Chikatamarla}\ \emph {et~al.}(2006)\citenamefont
  {Chikatamarla}, \citenamefont {Ansumali},\ and\ \citenamefont
  {Karlin}}]{chikatamarla2006entropic}%
  \BibitemOpen
  \bibfield  {author} {\bibinfo {author} {\bibfnamefont {S.~S.}~\bibnamefont
  {Chikatamarla}}, \bibinfo {author} {\bibfnamefont {S.}~\bibnamefont
  {Ansumali}}, \ and\ \bibinfo {author} {\bibfnamefont {I.~V.}~\bibnamefont
  {Karlin}},\ }\href@noop {} {\bibfield  {journal} {\bibinfo  {journal}
  {Physical review letters}\ }\textbf {\bibinfo {volume} {97}},\ \bibinfo
  {pages} {010201} (\bibinfo {year} {2006})}\BibitemShut {NoStop}%
\bibitem [{\citenamefont {Dorschner}\ \emph {et~al.}(2016)\citenamefont
  {Dorschner}, \citenamefont {B{\"o}sch}, \citenamefont {Chikatamarla},
  \citenamefont {Boulouchos},\ and\ \citenamefont
  {Karlin}}]{dorschner2016entropic}%
  \BibitemOpen
  \bibfield  {author} {\bibinfo {author} {\bibfnamefont {B.}~\bibnamefont
  {Dorschner}}, \bibinfo {author} {\bibfnamefont {F.}~\bibnamefont
  {B{\"o}sch}}, \bibinfo {author} {\bibfnamefont {S.~S.}\ \bibnamefont
  {Chikatamarla}}, \bibinfo {author} {\bibfnamefont {K.}~\bibnamefont
  {Boulouchos}}, \ and\ \bibinfo {author} {\bibfnamefont {I.~V.}\ \bibnamefont
  {Karlin}},\ }\href@noop {} {\bibfield  {journal} {\bibinfo  {journal}
  {Journal of Fluid Mechanics}\ }\textbf {\bibinfo {volume} {801}},\ \bibinfo
  {pages} {623} (\bibinfo {year} {2016})}\BibitemShut {NoStop}%
\bibitem [{\citenamefont {Ansumali}\ \emph {et~al.}(2003)\citenamefont
  {Ansumali}, \citenamefont {Karlin},\ and\ \citenamefont
  {{\"O}ttinger}}]{ansumali2003minimal}%
  \BibitemOpen
  \bibfield  {author} {\bibinfo {author} {\bibfnamefont {S.}~\bibnamefont
  {Ansumali}}, \bibinfo {author} {\bibfnamefont {I.~V.}\ \bibnamefont
  {Karlin}}, \ and\ \bibinfo {author} {\bibfnamefont {H.~C.}\ \bibnamefont
  {{\"O}ttinger}},\ }\href@noop {} {\bibfield  {journal} {\bibinfo  {journal}
  {EPL (Europhysics Letters)}\ }\textbf {\bibinfo {volume} {63}},\ \bibinfo
  {pages} {798} (\bibinfo {year} {2003})}\BibitemShut {NoStop}%
\bibitem [{\citenamefont {Kupershtokh}\ \emph {et~al.}(2009)\citenamefont
  {Kupershtokh}, \citenamefont {Medvedev},\ and\ \citenamefont
  {Karpov}}]{kupershtokh2009equations}%
  \BibitemOpen
  \bibfield  {author} {\bibinfo {author} {\bibfnamefont {A.}~\bibnamefont
  {Kupershtokh}}, \bibinfo {author} {\bibfnamefont {D.}~\bibnamefont
  {Medvedev}}, \ and\ \bibinfo {author} {\bibfnamefont {D.}~\bibnamefont
  {Karpov}},\ }\href@noop {} {\bibfield  {journal} {\bibinfo  {journal}
  {Computers \& Mathematics with Applications}\ }\textbf {\bibinfo {volume}
  {58}},\ \bibinfo {pages} {965} (\bibinfo {year} {2009})}\BibitemShut
  {NoStop}%
\bibitem [{\citenamefont {Quarteroni}\ and\ \citenamefont
  {Valli}(2008)}]{quarteroni2008numerical}%
  \BibitemOpen
  \bibfield  {author} {\bibinfo {author} {\bibfnamefont {A.}~\bibnamefont
  {Quarteroni}}\ and\ \bibinfo {author} {\bibfnamefont {A.}~\bibnamefont
  {Valli}},\ }\href@noop {} {\emph {\bibinfo {title} {Numerical approximation
  of partial differential equations}}},\ Vol.~\bibinfo {volume} {23}\ (\bibinfo
   {publisher} {Springer Science \& Business Media},\ \bibinfo {year}
  {2008})\BibitemShut {NoStop}%
\bibitem [{\citenamefont {Ga{\~n}{\'a}n-Calvo}(2004)}]{ganan2004general}%
  \BibitemOpen
  \bibfield  {author} {\bibinfo {author} {\bibfnamefont {A.~M.}\ \bibnamefont
  {Ga{\~n}{\'a}n-Calvo}},\ }\href@noop {} {\bibfield  {journal} {\bibinfo
  {journal} {Journal of fluid mechanics}\ }\textbf {\bibinfo {volume} {507}},\
  \bibinfo {pages} {203} (\bibinfo {year} {2004})}\BibitemShut {NoStop}%
\bibitem [{\citenamefont {McAllister}(1990)}]{mcallister1990conductor}%
  \BibitemOpen
  \bibfield  {author} {\bibinfo {author} {\bibfnamefont {I.}~\bibnamefont
  {McAllister}},\ }\href@noop {} {\bibfield  {journal} {\bibinfo  {journal}
  {Journal of Physics D: Applied Physics}\ }\textbf {\bibinfo {volume} {23}},\
  \bibinfo {pages} {359} (\bibinfo {year} {1990})}\BibitemShut {NoStop}%
\bibitem [{\citenamefont {Spencer}\ \emph {et~al.}(2010)\citenamefont
  {Spencer}, \citenamefont {Halliday},\ and\ \citenamefont
  {Care}}]{spencer2010lattice}%
  \BibitemOpen
  \bibfield  {author} {\bibinfo {author} {\bibfnamefont {T.~J.}~\bibnamefont
  {Spencer}}, \bibinfo {author} {\bibfnamefont {I.}~\bibnamefont {Halliday}}, \
  and\ \bibinfo {author} {\bibfnamefont {C.~M.}~\bibnamefont {Care}},\ }\href@noop
  {} {\bibfield  {journal} {\bibinfo  {journal} {Physical Review E}\ }\textbf
  {\bibinfo {volume} {82}},\ \bibinfo {pages} {066701} (\bibinfo {year}
  {2010})}\BibitemShut {NoStop}%
\bibitem [{\citenamefont {Yarin}\ \emph
  {et~al.}(2001{\natexlab{b}})\citenamefont {Yarin}, \citenamefont
  {Koombhongse},\ and\ \citenamefont {Reneker}}]{yarin2001taylor}%
  \BibitemOpen
  \bibfield  {author} {\bibinfo {author} {\bibfnamefont {A.~L.}\ \bibnamefont
  {Yarin}}, \bibinfo {author} {\bibfnamefont {S.}~\bibnamefont {Koombhongse}},
  \ and\ \bibinfo {author} {\bibfnamefont {D.~H.}\ \bibnamefont {Reneker}},\
  }\href@noop {} {\bibfield  {journal} {\bibinfo  {journal} {Journal of applied
  physics}\ }\textbf {\bibinfo {volume} {90}},\ \bibinfo {pages} {4836}
  (\bibinfo {year} {2001}{\natexlab{b}})}\BibitemShut {NoStop}%
\bibitem [{\citenamefont {Chikatamarla}\ \emph {et~al.}(2015)\citenamefont
  {Chikatamarla}, \citenamefont {Karlin} \emph
  {et~al.}}]{chikatamarla2015entropic}%
  \BibitemOpen
  \bibfield  {author} {\bibinfo {author} {\bibinfo {author} {\bibfnamefont {A.}\ \bibnamefont
  {Mazloomi M}},\bibfnamefont {S.~S.}~\bibnamefont
  {Chikatamarla}}, \bibinfo {author} {\bibfnamefont {I.~V.}~\bibnamefont
  {Karlin}},\ }\href@noop {} {\bibfield  {journal} {\bibinfo
  {journal} {Physical review letters}\ }\textbf {\bibinfo {volume} {114}},\
  \bibinfo {pages} {174502} (\bibinfo {year} {2015})}\BibitemShut {NoStop}%
\bibitem [{\citenamefont {Huang}\ \emph {et~al.}(2015)\citenamefont {Huang},
  \citenamefont {Sukop},\ and\ \citenamefont {Lu}}]{huang2015multiphase}%
  \BibitemOpen
  \bibfield  {author} {\bibinfo {author} {\bibfnamefont {H.}~\bibnamefont
  {Huang}}, \bibinfo {author} {\bibfnamefont {M.}~\bibnamefont {Sukop}}, \ and\
  \bibinfo {author} {\bibfnamefont {X.}~\bibnamefont {Lu}},\ }\href@noop {}
  {\emph {\bibinfo {title} {Multiphase lattice Boltzmann methods: Theory and
  application}}}\ (\bibinfo  {publisher} {John Wiley \& Sons},\ \bibinfo {year}
  {2015})\BibitemShut {NoStop}%
\bibitem [{\citenamefont {Falcucci}\ \emph {et~al.}(2010)\citenamefont
  {Falcucci}, \citenamefont {Ubertini},\ and\ \citenamefont
  {Succi}}]{falcucci2010lattice}%
  \BibitemOpen
  \bibfield  {author} {\bibinfo {author} {\bibfnamefont {G.}~\bibnamefont
  {Falcucci}}, \bibinfo {author} {\bibfnamefont {S.}~\bibnamefont {Ubertini}},
  \ and\ \bibinfo {author} {\bibfnamefont {S.}~\bibnamefont {Succi}},\
  }\href@noop {} {\bibfield  {journal} {\bibinfo  {journal} {Soft Matter}\
  }\textbf {\bibinfo {volume} {6}},\ \bibinfo {pages} {4357} (\bibinfo {year}
  {2010})}\BibitemShut {NoStop}%
\bibitem [{\citenamefont {Shan}\ and\ \citenamefont
  {Chen}(1993)}]{shan1993lattice}%
  \BibitemOpen
  \bibfield  {author} {\bibinfo {author} {\bibfnamefont {X.}~\bibnamefont
  {Shan}}\ and\ \bibinfo {author} {\bibfnamefont {H.}~\bibnamefont {Chen}},\
  }\href@noop {} {\bibfield  {journal} {\bibinfo  {journal} {Physical Review
  E}\ }\textbf {\bibinfo {volume} {47}},\ \bibinfo {pages} {1815} (\bibinfo
  {year} {1993})}\BibitemShut {NoStop}%
\bibitem [{\citenamefont {Pontrelli}\ \emph {et~al.}(2009)\citenamefont
  {Pontrelli}, \citenamefont {Ubertini},\ and\ \citenamefont
  {Succi}}]{pontrelli2009unstructured}%
  \BibitemOpen
  \bibfield  {author} {\bibinfo {author} {\bibfnamefont {G.}~\bibnamefont
  {Pontrelli}}, \bibinfo {author} {\bibfnamefont {S.}~\bibnamefont {Ubertini}},
  \ and\ \bibinfo {author} {\bibfnamefont {S.}~\bibnamefont {Succi}},\
  }\href@noop {} {\bibfield  {journal} {\bibinfo  {journal} {Journal of
  Statistical Mechanics: Theory and Experiment}\ }\textbf {\bibinfo {volume}
  {2009}},\ \bibinfo {pages} {P06005} (\bibinfo {year} {2009})}\BibitemShut
  {NoStop}%
\bibitem [{\citenamefont {Gabbanelli}\ \emph {et~al.}(2005)\citenamefont
  {Gabbanelli}, \citenamefont {Drazer},\ and\ \citenamefont
  {Koplik}}]{gabbanelli2005lattice}%
  \BibitemOpen
  \bibfield  {author} {\bibinfo {author} {\bibfnamefont {S.}~\bibnamefont
  {Gabbanelli}}, \bibinfo {author} {\bibfnamefont {G.}~\bibnamefont {Drazer}},
  \ and\ \bibinfo {author} {\bibfnamefont {J.}~\bibnamefont {Koplik}},\
  }\href@noop {} {\bibfield  {journal} {\bibinfo  {journal} {Physical review
  E}\ }\textbf {\bibinfo {volume} {72}},\ \bibinfo {pages} {046312} (\bibinfo
  {year} {2005})}\BibitemShut {NoStop}%
\bibitem [{\citenamefont {Aharonov}\ and\ \citenamefont
  {Rothman}(1993)}]{aharonov1993non}%
  \BibitemOpen
  \bibfield  {author} {\bibinfo {author} {\bibfnamefont {E.}~\bibnamefont
  {Aharonov}}\ and\ \bibinfo {author} {\bibfnamefont {D.~H.}\ \bibnamefont
  {Rothman}},\ }\href@noop {} {\bibfield  {journal} {\bibinfo  {journal}
  {Geophysical Research Letters}\ }\textbf {\bibinfo {volume} {20}},\ \bibinfo
  {pages} {679} (\bibinfo {year} {1993})}\BibitemShut {NoStop}%
\bibitem [{\citenamefont {Ouared}\ and\ \citenamefont
  {Chopard}(2005)}]{ouared2005lattice}%
  \BibitemOpen
  \bibfield  {author} {\bibinfo {author} {\bibfnamefont {R.}~\bibnamefont
  {Ouared}}\ and\ \bibinfo {author} {\bibfnamefont {B.}~\bibnamefont
  {Chopard}},\ }\href@noop {} {\bibfield  {journal} {\bibinfo  {journal}
  {Journal of statistical physics}\ }\textbf {\bibinfo {volume} {121}},\
  \bibinfo {pages} {209} (\bibinfo {year} {2005})}\BibitemShut {NoStop}%
\bibitem [{\citenamefont {Agarwal}\ \emph {et~al.}(2016)\citenamefont
  {Agarwal}, \citenamefont {Burgard}, \citenamefont {Greiner},\ and\
  \citenamefont {Wendorff}}]{agarwal2016electrospinning}%
  \BibitemOpen
  \bibfield  {author} {\bibinfo {author} {\bibfnamefont {S.}~\bibnamefont
  {Agarwal}}, \bibinfo {author} {\bibfnamefont {M.}~\bibnamefont {Burgard}},
  \bibinfo {author} {\bibfnamefont {A.}~\bibnamefont {Greiner}}, \ and\
  \bibinfo {author} {\bibfnamefont {J.}~\bibnamefont {Wendorff}},\ }\href@noop
  {} {\emph {\bibinfo {title} {Electrospinning: A Practical Guide to
  Nanofibers}}}\ (\bibinfo  {publisher} {Walter de Gruyter GmbH \& Co KG},\
  \bibinfo {year} {2016})\BibitemShut {NoStop}%
\bibitem [{\citenamefont {Johnson}(2016)}]{johnson2016handbook}%
  \BibitemOpen
  \bibfield  {author} {\bibinfo {author} {\bibfnamefont {R.~W.}\ \bibnamefont
  {Johnson}},\ }\href@noop {} {\emph {\bibinfo {title} {Handbook of fluid
  dynamics}}}\ (\bibinfo  {publisher} {Crc Press},\ \bibinfo {year}
  {2016})\BibitemShut {NoStop}%
\bibitem [{\citenamefont {Montinaro}\ \emph {et~al.}(2015)\citenamefont
  {Montinaro}, \citenamefont {Fasano}, \citenamefont {Moffa}, \citenamefont
  {Camposeo}, \citenamefont {Persano}, \citenamefont {Lauricella},
  \citenamefont {Succi},\ and\ \citenamefont {Pisignano}}]{montinaro2015sub}%
  \BibitemOpen
  \bibfield  {author} {\bibinfo {author} {\bibfnamefont {M.}~\bibnamefont
  {Montinaro}}, \bibinfo {author} {\bibfnamefont {V.}~\bibnamefont {Fasano}},
  \bibinfo {author} {\bibfnamefont {M.}~\bibnamefont {Moffa}}, \bibinfo
  {author} {\bibfnamefont {A.}~\bibnamefont {Camposeo}}, \bibinfo {author}
  {\bibfnamefont {L.}~\bibnamefont {Persano}}, \bibinfo {author} {\bibfnamefont
  {M.}~\bibnamefont {Lauricella}}, \bibinfo {author} {\bibfnamefont
  {S.}~\bibnamefont {Succi}}, \ and\ \bibinfo {author} {\bibfnamefont
  {D.}~\bibnamefont {Pisignano}},\ }\href@noop {} {\bibfield  {journal}
  {\bibinfo  {journal} {Soft matter}\ }\textbf {\bibinfo {volume} {11}},\
  \bibinfo {pages} {3424} (\bibinfo {year} {2015})}\BibitemShut {NoStop}%
\bibitem [{\citenamefont {Lauricella}\ \emph
  {et~al.}(2015{\natexlab{b}})\citenamefont {Lauricella}, \citenamefont
  {Pontrelli}, \citenamefont {Pisignano},\ and\ \citenamefont
  {Succi}}]{lauricella2015nonlinear}%
  \BibitemOpen
  \bibfield  {author} {\bibinfo {author} {\bibfnamefont {M.}~\bibnamefont
  {Lauricella}}, \bibinfo {author} {\bibfnamefont {G.}~\bibnamefont
  {Pontrelli}}, \bibinfo {author} {\bibfnamefont {D.}~\bibnamefont
  {Pisignano}}, \ and\ \bibinfo {author} {\bibfnamefont {S.}~\bibnamefont
  {Succi}},\ }\href@noop {} {\bibfield  {journal} {\bibinfo  {journal}
  {Molecular Physics}\ }\textbf {\bibinfo {volume} {113}},\ \bibinfo {pages}
  {2435} (\bibinfo {year} {2015}{\natexlab{b}})}\BibitemShut {NoStop}%
\bibitem [{\citenamefont {Lauricella}\ \emph
  {et~al.}(2015{\natexlab{c}})\citenamefont {Lauricella}, \citenamefont
  {Pontrelli}, \citenamefont {Coluzza}, \citenamefont {Pisignano},\ and\
  \citenamefont {Succi}}]{lauricella2015different}%
  \BibitemOpen
  \bibfield  {author} {\bibinfo {author} {\bibfnamefont {M.}~\bibnamefont
  {Lauricella}}, \bibinfo {author} {\bibfnamefont {G.}~\bibnamefont
  {Pontrelli}}, \bibinfo {author} {\bibfnamefont {I.}~\bibnamefont {Coluzza}},
  \bibinfo {author} {\bibfnamefont {D.}~\bibnamefont {Pisignano}}, \ and\
  \bibinfo {author} {\bibfnamefont {S.}~\bibnamefont {Succi}},\ }\href@noop {}
  {\bibfield  {journal} {\bibinfo  {journal} {Mechanics Research
  Communications}\ }\textbf {\bibinfo {volume} {69}},\ \bibinfo {pages} {97}
  (\bibinfo {year} {2015}{\natexlab{c}})}\BibitemShut {NoStop}%
\bibitem [{\citenamefont {Lauricella}\ \emph {et~al.}(2017)\citenamefont
  {Lauricella}, \citenamefont {Cipolletta}, \citenamefont {Pontrelli},
  \citenamefont {Pisignano},\ and\ \citenamefont
  {Succi}}]{lauricella2017effects}%
  \BibitemOpen
  \bibfield  {author} {\bibinfo {author} {\bibfnamefont {M.}~\bibnamefont
  {Lauricella}}, \bibinfo {author} {\bibfnamefont {F.}~\bibnamefont
  {Cipolletta}}, \bibinfo {author} {\bibfnamefont {G.}~\bibnamefont
  {Pontrelli}}, \bibinfo {author} {\bibfnamefont {D.}~\bibnamefont
  {Pisignano}}, \ and\ \bibinfo {author} {\bibfnamefont {S.}~\bibnamefont
  {Succi}},\ }\href@noop {} {\bibfield  {journal} {\bibinfo  {journal} {Physics
  of Fluids}\ }\textbf {\bibinfo {volume} {29}},\ \bibinfo {pages} {082003}
  (\bibinfo {year} {2017})}\BibitemShut {NoStop}%
\bibitem [{\citenamefont {Greenfeld}\ \emph {et~al.}(2011)\citenamefont
  {Greenfeld}, \citenamefont {Arinstein}, \citenamefont {Fezzaa}, \citenamefont
  {Rafailovich},\ and\ \citenamefont {Zussman}}]{greenfeld2011polymer}%
  \BibitemOpen
  \bibfield  {author} {\bibinfo {author} {\bibfnamefont {I.}~\bibnamefont
  {Greenfeld}}, \bibinfo {author} {\bibfnamefont {A.}~\bibnamefont
  {Arinstein}}, \bibinfo {author} {\bibfnamefont {K.}~\bibnamefont {Fezzaa}},
  \bibinfo {author} {\bibfnamefont {M.~H.}\ \bibnamefont {Rafailovich}}, \ and\
  \bibinfo {author} {\bibfnamefont {E.}~\bibnamefont {Zussman}},\ }\href@noop
  {} {\bibfield  {journal} {\bibinfo  {journal} {Physical Review E}\ }\textbf
  {\bibinfo {volume} {84}},\ \bibinfo {pages} {041806} (\bibinfo {year}
  {2011})}\BibitemShut {NoStop}%
\bibitem [{\citenamefont {Reznik}\ and\ \citenamefont
  {Zussman}(2010)}]{reznik2010capillary}%
  \BibitemOpen
  \bibfield  {author} {\bibinfo {author} {\bibfnamefont {S.~N.}~\bibnamefont
  {Reznik}}\ and\ \bibinfo {author} {\bibfnamefont {E.}~\bibnamefont
  {Zussman}},\ }\href@noop {} {\bibfield  {journal} {\bibinfo  {journal}
  {Physical Review E}\ }\textbf {\bibinfo {volume} {81}},\ \bibinfo {pages}
  {026313} (\bibinfo {year} {2010})}\BibitemShut {NoStop}%
\bibitem [{\citenamefont {Higuera}(2003)}]{higuera2003flow}%
  \BibitemOpen
  \bibfield  {author} {\bibinfo {author} {\bibfnamefont {F.}~\bibnamefont
  {Higuera}},\ }\href@noop {} {\bibfield  {journal} {\bibinfo  {journal}
  {Journal of Fluid Mechanics}\ }\textbf {\bibinfo {volume} {484}},\ \bibinfo
  {pages} {303} (\bibinfo {year} {2003})}\BibitemShut {NoStop}%
\bibitem [{\citenamefont {Hohman}\ \emph
  {et~al.}(2001{\natexlab{b}})\citenamefont {Hohman}, \citenamefont {Shin},
  \citenamefont {Rutledge},\ and\ \citenamefont
  {Brenner}}]{hohman2001electrospinning2}%
  \BibitemOpen
  \bibfield  {author} {\bibinfo {author} {\bibfnamefont {M.~M.}\ \bibnamefont
  {Hohman}}, \bibinfo {author} {\bibfnamefont {M.}~\bibnamefont {Shin}},
  \bibinfo {author} {\bibfnamefont {G.}~\bibnamefont {Rutledge}}, \ and\
  \bibinfo {author} {\bibfnamefont {M.~P.}\ \bibnamefont {Brenner}},\
  }\href@noop {} {\bibfield  {journal} {\bibinfo  {journal} {Physics of
  fluids}\ }\textbf {\bibinfo {volume} {13}},\ \bibinfo {pages} {2221}
  (\bibinfo {year} {2001}{\natexlab{b}})}\BibitemShut {NoStop}%
\bibitem [{\citenamefont {Hartman}\ \emph {et~al.}(1999)\citenamefont
  {Hartman}, \citenamefont {Brunner}, \citenamefont {Camelot}, \citenamefont
  {Marijnissen},\ and\ \citenamefont
  {Scarlett}}]{hartman1999electrohydrodynamic}%
  \BibitemOpen
  \bibfield  {author} {\bibinfo {author} {\bibfnamefont {R.~P.~A.}\
  \bibnamefont {Hartman}}, \bibinfo {author} {\bibfnamefont {D.}~\bibnamefont
  {Brunner}}, \bibinfo {author} {\bibfnamefont {D.}~\bibnamefont {Camelot}},
  \bibinfo {author} {\bibfnamefont {J.}~\bibnamefont {Marijnissen}}, \ and\
  \bibinfo {author} {\bibfnamefont {B.}~\bibnamefont {Scarlett}},\ }\href@noop
  {} {\bibfield  {journal} {\bibinfo  {journal} {Journal of Aerosol science}\
  }\textbf {\bibinfo {volume} {30}},\ \bibinfo {pages} {823} (\bibinfo {year}
  {1999})}\BibitemShut {NoStop}%
\bibitem [{\citenamefont {Cherney}(1999)}]{cherney1999structure}%
  \BibitemOpen
  \bibfield  {author} {\bibinfo {author} {\bibfnamefont {L.~T.}\ \bibnamefont
  {Cherney}},\ }\href@noop {} {\bibfield  {journal} {\bibinfo  {journal}
  {Journal of Fluid Mechanics}\ }\textbf {\bibinfo {volume} {378}},\ \bibinfo
  {pages} {167} (\bibinfo {year} {1999})}\BibitemShut {NoStop}%
\bibitem [{\citenamefont {Arinstein}(2017)}]{arinstein2017electrospun}%
  \BibitemOpen
  \bibfield  {author} {\bibinfo {author} {\bibfnamefont {A.}~\bibnamefont
  {Arinstein}},\ }\href@noop {} {\emph {\bibinfo {title} {Electrospun Polymer
  Nanofibers: A physicist's point of view}}}\ (\bibinfo  {publisher} {CRC
  Press},\ \bibinfo {year} {2017})\BibitemShut {NoStop}%
\bibitem [{\citenamefont {Thompson}\ \emph {et~al.}(2007)\citenamefont
  {Thompson}, \citenamefont {Chase}, \citenamefont {Yarin},\ and\ \citenamefont
  {Reneker}}]{thompson2007effects}%
  \BibitemOpen
  \bibfield  {author} {\bibinfo {author} {\bibfnamefont {C.}~\bibnamefont
  {Thompson}}, \bibinfo {author} {\bibfnamefont {G.~G.}\ \bibnamefont {Chase}},
  \bibinfo {author} {\bibfnamefont {A.}~\bibnamefont {Yarin}}, \ and\ \bibinfo
  {author} {\bibfnamefont {D.}~\bibnamefont {Reneker}},\ }\href@noop {}
  {\bibfield  {journal} {\bibinfo  {journal} {Polymer}\ }\textbf {\bibinfo
  {volume} {48}},\ \bibinfo {pages} {6913} (\bibinfo {year}
  {2007})}\BibitemShut {NoStop}%
\bibitem [{\citenamefont {Christanti}\ and\ \citenamefont
  {Walker}(2001)}]{christanti2001surface}%
  \BibitemOpen
  \bibfield  {author} {\bibinfo {author} {\bibfnamefont {Y.}~\bibnamefont
  {Christanti}}\ and\ \bibinfo {author} {\bibfnamefont {L.~M.}\ \bibnamefont
  {Walker}},\ }\href@noop {} {\bibfield  {journal} {\bibinfo  {journal}
  {Journal of Non-Newtonian Fluid Mechanics}\ }\textbf {\bibinfo {volume}
  {100}},\ \bibinfo {pages} {9} (\bibinfo {year} {2001})}\BibitemShut {NoStop}%
\bibitem [{\citenamefont {Cabral}\ and\ \citenamefont
  {Leedom}(1993)}]{cabral1993imaging}%
  \BibitemOpen
  \bibfield  {author} {\bibinfo {author} {\bibfnamefont {B.}~\bibnamefont
  {Cabral}}\ and\ \bibinfo {author} {\bibfnamefont {L.~C.}\ \bibnamefont
  {Leedom}},\ }in\ \href@noop {} {\emph {\bibinfo {booktitle} {Proceedings of
  the 20th annual conference on Computer graphics and interactive
  techniques}}}\ (\bibinfo {organization} {ACM},\ \bibinfo {year} {1993})\ pp.\
  \bibinfo {pages} {263--270}\BibitemShut {NoStop}%
\end{thebibliography}
\end{document}